\documentclass[structabstract]{aa}  

\usepackage{graphicx}
\usepackage{txfonts}
\usepackage{natbib}
\usepackage{color}
\usepackage{threeparttable}

\newcommand{\htwoo}{H$_{2}$O}

\newcommand{\msol}{$\rm M_{\odot}$}

\newcommand{\mdot}{$\rm \dot{M}$}
\newcommand{\mstar}{$\rm M_{\star}$}
\newcommand{\lsol}{$\rm L_{\odot}$}
{
\newcommand{\lstar}{$\rm L_{\star}$}
\newcommand{\rsol}{$\rm R_{\odot}$}
\newcommand{\rstar}{$\rm R_{\star}$}
\newcommand{\tkin}{$T_{\rm kin}$}
\newcommand{\tdust}{$T_{\rm dust}$}
\newcommand{\ngas}{$n_{\rm H_{2}}$}

\newcommand{\teff}{$T_{\rm eff}$}
\newcommand{\kms}{km$\,$s$^{-1}$}

\newcommand{\mearth}{$M_{\oplus}$}

\newcommand{\chgas}{[C]/[H]$_{\rm gas}$}
\newcommand{\ohgas}{[O]/[H]$_{\rm gas}$}

\newcommand{\gdrat}{$\Delta_{\rm gas/dust}$}

\newcommand{\cplus}{C$^{+}$}
\newcommand{\catom}{C$^{0}$}
\newcommand{\ci}{[\ion{C}{i}]}
\newcommand{\cii}{[\ion{C}{ii}]}
\newcommand{\oi}{[\ion{O}{i}]}
\newcommand{\cosixfive}{CO~$6$--$5$}
\newcommand{\cionezero}{[\ion{C}{i}]~$1$--$0$}
\newcommand{\citwoone}{[\ion{C}{i}]~$2$--$1$}

\newcommand{\champp}{CHAMP$^{+}$}

\begin{document}
\title{Volatile carbon locking and release in protoplanetary disks\thanks{Based on observations collected at the European Organisation for Astronomical Research in the Southern Hemisphere under ESO programmes 093.C-0926, 093.F-0015, 077.D-0092, 084.A-9016, and 085.A-9027.}}
\subtitle{A study of TW~Hya and HD~100546}

   \author{
   M. Kama\inst{1}
          \and
          S. Bruderer\inst{2}
          \and
          E.F. van Dishoeck\inst{1}
          \and
          M. Hogerheijde\inst{1}
          \and
          C.P. Folsom\inst{3,4}
          \and
          A. Miotello\inst{1}
          \and
          D. Fedele\inst{2}
          \and
          A. Belloche\inst{5}
          \and
          R. G\"{u}sten\inst{5}
          \and
          F. Wyrowski\inst{5}
          }

\institute{
		Leiden Observatory, Leiden University, P.O. Box 9513, NL-2300 RA, Leiden, The Netherlands, \email{mkama@strw.leidenuniv.nl}
		\and
		INAF-Osservatorio Astrofisico di Arcetri, Largo E. Fermi 5, 50125, Firenze, Italy
		\and
		Universit\'{e} de Grenoble Alpes, IPAG, F-38000 Grenoble, France
		\and
		CNRS, IPAG, F-38000 Grenoble, France
		\and
		Max-Planck-Institut f\"{u}r Radioastronomie, Auf dem H\"{u}gel 69, 53121, Bonn, Germany
             }

   \date{}

  \abstract
  {}  
   {The composition of planetary solids and gases is largely rooted in the processing of volatile elements in protoplanetary disks. To shed light on the key processes, we carry out a comparative analysis of the gas-phase carbon abundance in two systems with a similar age and disk mass, but different central stars: HD~100546 and TW~Hya.}
   {We combine our recent detections of C$^{0}$ in these disks with observations of other carbon reservoirs (CO, C$^{+}$, C$_{2}$H) and gas mass and warm gas tracers (HD, O$^{0}$), as well as spatially resolved ALMA observations and the spectral energy distribution. The disks are modelled with the DALI 2D physical-chemical code. Stellar abundances for HD~100546 are derived from archival spectra.}
   {Upper limits on HD emission from HD~100546 place an upper limit on the total disk mass of $\leq0.1\,$\msol. The gas-phase carbon abundance in the atmosphere of this warm Herbig disk is at most moderately depleted compared to the interstellar medium, with [C]/[H]$_{\rm gas}=(0.1-1.5)\times 10^{-4}$. HD~100546 itself is a $\lambda\,$Bo\"{o}tis star, with solar abundances of C and O but a strong depletion of rock-forming elements. In the gas of the T~Tauri disk TW~Hya, both C and O are strongly underabundant, with [C]/[H]$_{\rm gas}=(0.2-5.0)\times 10^{-6}$ and C/O${>}1$. We discuss evidence that the gas-phase C and O abundances are high in the warm inner regions of both disks. Our analytical model, including vertical mixing and a grain size distribution, reproduces the observed [C]/[H]$_{\rm gas}$ in the outer disk of TW~Hya and allows to make predictions for other systems.}
   {}
   \keywords{Astrochemistry; Protoplanetary disks}
   \maketitle
   
\section{Introduction}

Several protoplanetary disks have been found to be depleted of gas-phase CO. This is generally understood to result from the freezeout of molecules in the midplane and from CO photodissociation in the disk atmosphere \citep{Dutreyetal1996, Dutreyetal1997, Dutreyetal2003, vanZadelhoffetal2001,Chapillonetal2008, Chapillonetal2010}. Studies that account for these processes have suggested that the gas is genuinely underabundant in carbon in the disks around HD~100546 and TW~Hya \citep{Brudereretal2012, Berginetal2013, Favreetal2013, Duetal2015}. We revisit the gas-phase carbon abundance in these disks, including new detections of far-infrared \ci\ lines in the analysis, and present an analytical model for describing the volatile loss.

The solar elemental carbon abundance is [C]/[H]~$=2.69\times10^{-4}$ \citep{Asplundetal2009}, while the gas-phase abundance in the diffuse interstellar medium is \chgas~$=(1-2)\times10^{-4}$ \citep{Cardellietal1996, Parvathietal2012}. Several tens of percent of interstellar elemental carbon is locked in refractory solids. Processes affecting the gas and solid carbon budget in disks include the conversion of gas-phase CO into hydrocarbons \citep[e.g.][]{Aikawaetal1999, FuruyaAikawa2014,Berginetal2014}; the freeze out of CO and other carrier molecules \citep{Obergetal2011e}; the transformation of this icy reservoir into complex organic molecules in cold regions \citep{vanDishoeck2008, Walshetal2014a}; and the vertical transport and oxidation of carbonaceous solids in warm gas \citep{Leeetal2010}. In our models, we explicitly treat only the first two processes, but we argue that information about the rest can also be inferred if the carbon abundance in a disk atmosphere is determined.

The composition of a giant planet atmosphere is determined by the gas and dust in the formation or feeding zone of the planet; and by accreted icy planetesimals \citep{Pollacketal1996, Chabrieretal2014}. Formation by gravitational instability involves dust evaporation and leads to planets with roughly stellar atmospheric abundances. Core accretion, however, leads to an atmospheric composition determined by the local gas composition, with a potentially large contribution from ablated planetesimals. Determining the gas-solid budget of volatiles in disks is then important for understanding the potential diversity of giant planet atmospheres, but also for using statistics of planetary compositions as a test of planet population synthesis models \citep[e.g.][]{Pontoppidanetal2014PPVI, Thiabaudetal2015}. The gas-solid budget of carbon in a disk may even determine whether or not planets forming in it are habitable from a geophysical viewpoint \citep{Unterbornetal2014}.

This study is part of the project ``Disk Carbon and Oxygen'' (DISCO), aimed at studying the abundance and budget of key elements in planet-forming environments.

\section{Observations}\label{sec:obs}

\begin{figure}[!ht]
\includegraphics[clip=, width=1.0\columnwidth]{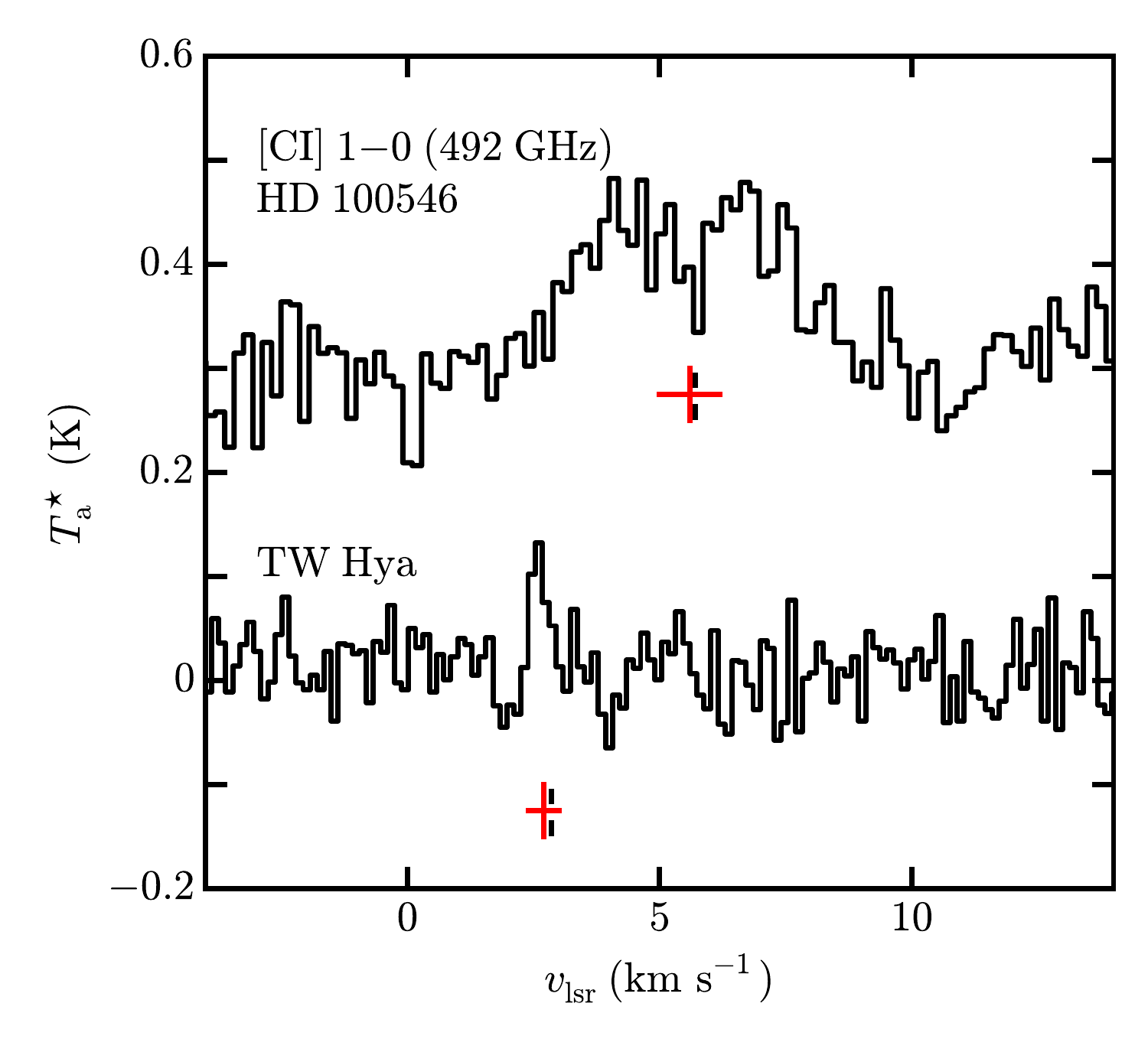}
\caption{The [CI]~$1-0$ line in HD~100546 (offset by $+0.3$~K) and TW~Hya. Horizontal bars mark the systemic velocities, $5.70$ and $2.86$~km$\cdot$s$^{-1}$. Red crosses show the best-fit Gaussian $v_{\rm lsr}$ with $3\,\sigma$ errorbars. The data are binned to $0.14$~km$\cdot$s$^{-1}$ per channel.}
\label{fig:cidetections}
\end{figure}

Deep integrations of the \ci, CO, C$_{2}$H and HCO$^{+}$ transitions, listed in Table~\ref{tab:newobs}, towards \object{TW~Hya} and \object{HD~100546} were obtained with the single-pixel FLASH \citep{Heymincketal2006} and $7$-pixel \champp\ (\citwoone, \cosixfive) receivers on APEX \citep{Gustenetal2006}. Observations of TW~Hya (ESO proposal E-093.C-0926A, PI M.Kama) totalled $9.1$~hours in September 2014, with a typical water vapour column of PWV~$\approx1.4\,$mm. HD~100546 (Max Planck proposal M0015\_93, PI E.F.van Dishoeck) was observed from April to July 2014, with PWV~$\approx0.75\,$mm. The backends were AFFTS (with a highest channel resolution of $0.18$~MHz or $0.11$~km$\,$s$^{-1}$ at $492$~GHz) and XFFTS (FLASH, $0.04$~MHz or $0.02$~\kms). Initial processing was done using the APECS software \citep{Mudersetal2006}. Inspection, baseline subtraction and averaging were done with GILDAS/CLASS\footnote{\texttt{http://www.iram.fr/IRAMFR/GILDAS}}. Telescope parameters were obtained from \citet{Gustenetal2006}. The kelvin to jansky conversion and main beam efficiency at 650~GHz (\champp-I) are $53$~Jy$\,$K$^{-1}$ and $0.56$, respectively. At 812~GHz (\champp-II), they are $70$~Jy$\,$K$^{-1}$ and $0.43$. At $491$~GHz (FLASH460), the conversion factor is $41$~Jy$\,$K$^{-1}$ and $\eta_{mb}=0.73$. The \ci\ detections shown in Fig.~\ref{fig:cidetections} and the \cosixfive\ detections were presented, but not analyzed in detail, as part of a large survey in \citet{Kamaetal2016a}.

\begin{table}[!ht]
\centering
\caption{New observations of TW~Hya and HD~100546 with APEX.}
\label{tab:newobs}
\begin{tabular}{ c c c c }
\hline
\hline
Transition								& $\nu$	& \multicolumn{2}{c}{$\int{T_{A}^{*}}dv$}	\\
									& (GHz)	& \multicolumn{2}{c}{(K$\,$km$\,$s$^{-1}$)}	\\
\hline
									&			& TW~Hya		& HD~100546	\\
\hline
\cionezero								& $492.161$	& $0.05\pm0.01$	& $0.49\pm0.05$	\\
\cosixfive								& $691.473$	& $1.0\pm0.1$		& $6.4\pm0.1$	\\
C$_{2}$H\,$4{-}3,\frac{9}{2}{-}\frac{7}{2},5{-}4$	& $349.338$	& $0.15\pm0.05$	& $\leq 0.03$	\\
C$_{2}$H\,$4{-}3,\frac{9}{2}{-}\frac{7}{2},4{-}3$	& $349.339$	& $0.12\pm0.05$	& $\leq 0.03$	\\
C$_{2}$H\,$4{-}3,\frac{7}{2}{-}\frac{5}{2},4{-}3$	& $349.399$	& $0.13\pm0.05$	& $\leq 0.03$	\\
C$_{2}$H\,$4{-}3,\frac{7}{2}{-}\frac{5}{2},3{-}2$	& $349.401$	& $0.10\pm0.05$	& $\leq 0.03$	\\
HCO$^{+}\,4$--$3$						& $356.734$	& $0.43\pm0.01$ 	& $0.14\pm0.02$	\\
\hline
\end{tabular}
\flushleft
\emph{Notes. }Uncertainties are at $1\,\sigma$ and upper limits at $3\,\sigma$ confidence. The quantum numbers for C$_{2}$H are $N, J, F$.
\end{table}

Supplementary archival optical spectra were used to perform a chemical abundance analysis of the photosphere of HD~100546. Spectra from FEROS \citep{Kauferetal1999} were extracted from the ESO Science Archive. FEROS is an echelle spectrograph mounted on the MPG/ESO 2.2m Telescope in La Silla. The instrument has a resolving power of R${\sim}48000$ and covers a wavelength range of $360$--$920\,$nm in $39$ spectral orders. The spectra were obtained in projects E-077.D-0092A, E-084.A-9016A, and E-085.A-9027B, in June 2006, March 2010, and May-June 2010, respectively. The reduced spectra extracted from the ESO Science Archive were continuum normalized by fitting a low order polynomial through carefully selected points. This was done for the individual spectral orders independently, then the normalized spectra were merged. The $10$ available observations were co-added to boost our final S/N, since no variability was found in the spectra outside of emission lines. The derivation of photospheric abundances from these data is described in Appendix~\ref{sec:hd546abuns}.

\section{The sources}\label{sec:sources}

Both of the targeted disks are relatively old ($\tau\sim 10\,$Myr) and have a similar dust mass (M$_{\rm dust}\sim 10^{-4}\,$\msol). The stellar effective temperatures and luminosities are very different, however -- $4110\,$K and $0.28\,$\lsol\ for TW~Hya and $10390\,$K and $36\,$\lsol\ for HD~100546 -- providing an opportunity to compare two similar, isolated disks with a very different average dust temperature. The properties of both sources are summarized in Table~\ref{tab:params}.

\emph{\textbf{HD~100546}} is a $(2.4\pm0.1)$~\msol\ disk-hosting star of spectral type B9V, with an estimated age of $\gtrsim10$~Myr \citep{vandenAnckeretal1997b}. The distance, measured with \emph{Hipparcos}, is $97\pm4\,$pc \citep{vanLeeuwen2007}. For its stellar properties and spectrum, including X-ray and UV, we adopt the values of \citet[][and references therein]{Brudereretal2012}. We note that the observed total luminosity of HD~100546, $36\,$\lsol, is larger than the $(25\pm7)\,$\lsol\ obtained from the photospheric abundance analysis in Appendix~\ref{sec:hd546abuns}, where the FUV data were not included.

The observable dust mass is $\approx10^{-4}$~\msol\ \citep{Henningetal1998, Wilneretal2003, PanicHogerheijde2009b, Muldersetal2011, Muldersetal2013}. Micron-sized grains are observed out to $1000\,$au \citep{Pantinetal2000, Augereauetal2001, Gradyetal2001, Ardilaetal2007}, while millimetre-sized grains are abundant within $230\,$au and the molecular gas, traced by CO isotopologs, extends to $390\,$au \citep{PanicHogerheijde2009b, Panicetal2010, Walshetal2014b}. The dust disk inclination is $44^{\circ}\pm3^{\circ}$. Large dust grains are depleted inside of $13\,$au, and again from $35$ to $150\,$au \citep[e.g.][]{Bouwmanetal2003, Walshetal2014b}, with a small inner disk from $0.26$ to $0.7$~au containing $3\times 10^{-10}$~\msol\ of dust \citep{Benistyetal2010, Panicetal2014}. The inner hole and outer gap are consistent with radial zones cleared partly of material by two substellar companions \citep{Walshetal2014b}.

\emph{\textbf{TW~Hya}} is considered to be a K7 spectral type pre-main sequence star with a mass of $0.8$~\msol, radius $1.04$~\rsol, \teff~$=4110$~K and a luminosity $0.28$~\lsol; although there is some debate regarding these parameters \citep[e.g.][]{VaccaSandell2011, Andrewsetal2012, Debesetal2013}. The distance is $(55\pm9)\,$pc \citep{vanLeeuwen2007} and the age $\sim8$--$10$~Myr \citep[e.g.][]{Hoffetal1998, Webbetal1999, delaReza2006, Debesetal2013}.

The disk gas mass was recently estimated from HD emission to be $\gtrsim 0.05$~\msol\ \citep{Gortietal2011, Berginetal2013}. An inner disk extends from $0.06$ to $0.5\,$au, followed by a radial gap in continuum opacity out to $4\,$au \citep{Calvetetal2002, Eisneretal2006, Hughesetal2007, Akesonetal2011, Andrewsetal2012}. The disk extends to $\gtrsim 280\,$au in scattered light, while the large grains are seen only out to $60\,$au in $870\,\mu$m emission and the CO gas is within $215\,$au \citep{Andrewsetal2012}. The inclination is $7^{\circ}\pm1^{\circ}$ \citep{Qietal2004, Hughesetal2011, Andrewsetal2012, Rosenfeldetal2012}. Scattered light imaging with HST at $0.5$ to $2.2\,\mu$m reveals a $30$\% intensity gap in the radial brightness distribution at $\sim80\,$au, consistent with a $6$ to $28$~\mearth\ companion \citep{Debesetal2013}. Another gap has recently been suggested at $\sim20\,$au \citep{Akiyamaetal2015}.

The accretion luminosity is estimated to be $0.03\,$\lsol, but estimates of the accretion rate vary from $4\times 10^{-10}$ to $4.8\times 10^{-9}$~\msol$\,$yr$^{-1}$ \citep{Muzerolleetal2000, Batalhaetal2002, Kastneretal2002, Debesetal2013}. The X-ray luminosity is $1.4\times 10^{30}$~erg$\,$s$^{-1}$, originating in a two-temperature plasma with a main peak at $T_{\rm X}=3.2\times 10^{6}$~K and a secondary at $\gtrsim 3.2\times 10^{7}$~K \citep{Kastneretal2002, Brickhouseetal2010}.

\section{Overview of the spectra}\label{sec:data}

The \cosixfive, \cionezero\ and HCO$^{+}$~$4$--$3$ transitions are detected towards both TW~Hya and HD~100546, while C$_{2}$H is detected only towards TW~Hya (Table~\ref{tab:newobs}). The \cionezero\ spectra are shown in Fig.~\ref{fig:cidetections}. The emission towards HD~100546 is double-peaked, with symmetric peaks for \ci\ and a slightly stronger blue- than redshifted peak for \cosixfive, consistent with previous CO observations \citep[e.g.][]{Panicetal2010} and with the inclination. Based on an extended scattered light halo and single-peaked \cii\ emission, a compact, tenuous envelope is thought to surround the HD~100546 disk \citep{Gradyetal2001, Brudereretal2012, Fedeleetal2013a}. The \cosixfive\ and \cionezero\ lines are double-peaked, indicating they are not contaminated by the envelope. As described in Appendix~\ref{sec:contaminationcheck}, we used a set of offset pointings to further verify that the \cosixfive\ and \cionezero\ transitions do not have a substantial envelope emission contribution. The line profiles towards TW~Hya are single-peaked and narrow, consistent with its low inclination. The \ci\ line peaks at $2.7\pm0.3\,$\kms\ ($3\,\sigma$), consistent with the canonical systemic velocity of $2.86\,$\kms. The C$_{2}$H line fluxes for TW~Hya are consistent with those from \citet{Kastneretal2014, Kastneretal2015}.

The detections of \cionezero\ represent the first unambiguous detections of atomic carbon in protoplanetary disks in the far-infrared. The line was detected previously towards DM~Tau by \citet{Tsukagoshietal2015} and \citet{Kamaetal2016a}, however it is single-peaked and narrower than the double-peaked CO profiles seen from that source, which suggests contamination by a compact residual envelope and requires further follow-up.

Our new data, listed in Table~\ref{tab:newobs}, were supplemented with continuum and spectral line data from the literature, listed in Appendix~\ref{apx:allobs}.

\section{Analysis}\label{sec:analysis}

To derive the gas-phase abundance of elemental carbon (\chgas), we have run source-specific model grids using the 2D physical-chemical code, \texttt{DALI} \citep{Brudereretal2009,Brudereretal2012,Brudereretal2013}. The code uses Monte Carlo radiative transfer to determine the dust temperature. The gas-grain chemistry in each grid cell is then solved time-dependently. For the grain surface chemistry, only hydrogenation is considered. The steady-state heating-cooling balance of the gas is determined at each time-step. We let the chemistry evolve to $10\,$Myr to match the ages of the two disks.

\begin{figure}[!ht]
\includegraphics[clip=, width=1.0\columnwidth]{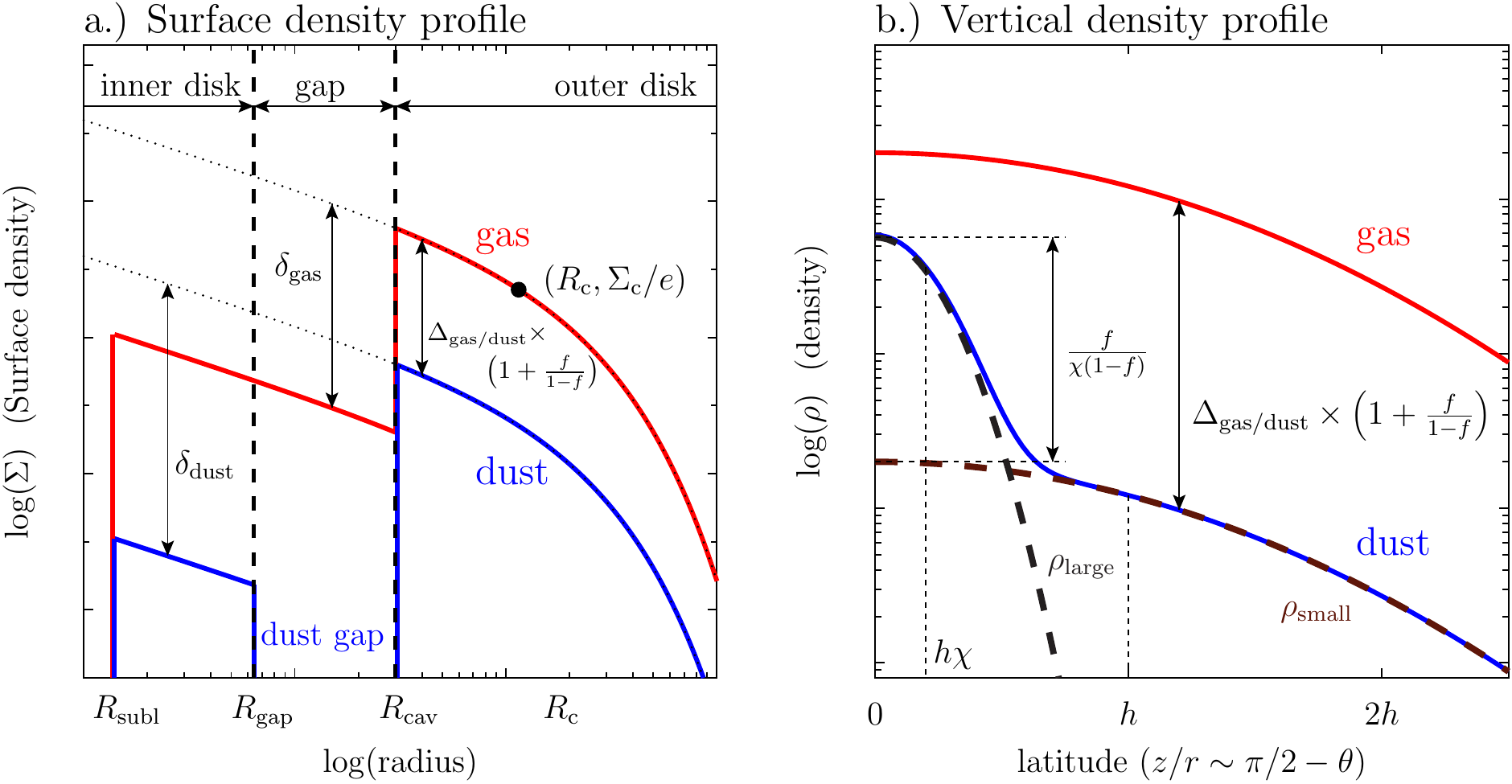}
\caption{The parametric disk structure used in DALI. See also Table~\ref{tab:params}.}
\label{fig:dalidisk}
\end{figure}

The disk structure is parameterized as shown in Fig.~\ref{fig:dalidisk}. We denote the gas-to-dust mass ratio as \gdrat. The dust consists of a small ($0.005$--$1\,\mu$m) and large ($0.005$--$1\,$mm) grain population. The surface density profile is a power law with an exponential taper:
\begin{equation}
\Sigma_{\rm gas} = \Sigma_{\rm c}\cdot \left( \frac{r}{R_{\rm c}} \right)^{-\gamma} \cdot \exp{\left[ - \left(\frac{r}{R_{\rm c}}\right)^{2-\gamma} \right]}.
\end{equation}

$\Sigma_{\rm gas}$ and $\Sigma_{\rm dust}$ extend from the dust sublimation radius $R_{\rm sub}$ to $R_{\rm out}$ and can be independently varied inside the radius $R_{\rm cav}$ with the multiplication factors $\delta_{\rm gas}$ and $\delta_{\rm dust}$. Dust is depleted entirely from $R_{\rm gap}$ to $R_{\rm cav}$. The scaleheight angle, $h$, at distance $r$ is given by $h(r)=h_{\rm c}\,(r/R_{\rm c})^{\psi}$. The scaleheight is then $H=h\cdot r$, and the vertical density distribution of the small grains is given by
\begin{equation}\label{eq:rhosmall}
\rho_{\rm dust,small} = \frac{(1-f)\,\Sigma_{dust}}{\sqrt{2\,\pi}\,r\,h} \times \exp{ \left[ -\frac{1}{2} \left( \frac{\pi/2 - \theta}{h} \right)^{2} \right] },
\end{equation}

where $f$ is the mass fraction of large grains, and $\theta$ is the opening angle from the midplane as viewed from the central star. The settling of large grains is prescribed as a fraction $\chi \in{(0,1]}$ of the scaleheight of the small grains, so the mass density of large grains is similar to Eq.~\ref{eq:rhosmall}, with $f$ replacing $(1-f)$ and $\chi\,h$ replacing $h$. The vertical distribution of gas is $\rho_{\rm gas}=\Delta_{\rm gas/dust}\times \rho_{\rm dust,small}\times [1+f/(1-f)$]. The term $f/(1-f)$ preserves the global \gdrat\ outside $R_{\rm cav}$.

\begin{table}[!ht]
\centering
\caption{Parameters for both sources. For definitions, see Fig.~\ref{fig:dalidisk}.}
\label{tab:params}
\begin{tabular}{ c c c }
\hline
\hline
Parameter			&	HD~100546				&	TW~Hya		\\
\hline
	&	\multicolumn{2}{c}{Star} \\
\hline
\lstar 				&	$36\,$\lsol			&	$0.28\,$\lsol		\\
\teff 					&	$10390\pm600\,$K		&	$4110\,$K		\\
\mstar 				&	$2.3\pm0.2\,$\msol		&	$0.74\,$\msol		\\
\rstar 				&	$1.5\pm0.3\,$\rsol		&	$1.05\,$\rsol				\\
$d$					&	$97\,$pc				&	$55\,$pc			\\
\hline
	&	\multicolumn{2}{c}{Disk} \\
\hline
Inner disk $R_{\rm subl}$--$R_{\rm gap}$	&	$0.25$--$4.0\,$au				&	$0.05$--$0.3\,$au 	\\
Inner hole $R_{\rm cav}$		&	$13\,$au						&	$4\,$au	\\
Inner hole $\Sigma$ scalings 	&	$\delta_{\rm gas}=10^{0}$--$10^{1}$		& $\delta_{\rm gas}\sim10^{-2}$	\\
							&	$\delta_{\rm dust}=10^{-5}$	& $\delta_{\rm dust}=10^{-2}$	\\
$\gamma$					&	$1.0$						&	$1.0$	\\
$\chi$, $f$	&	$0.8$, $0.85$				&	$0.2$, $0.99$	\\
$R_{\rm c}$					&	$50$~au						&	$35$~au	 \\
$\Sigma_{\rm c}$				&	$58$~g$\,$cm$^{-2}$		&	$30$~g$\,$cm$^{-2}$	\\
$h_{\rm c}$, $\psi$			&	$0.10$, $0.25$				&	$0.10$, $0.30$	\\
\gdrat						&	$10$--$300$					&	$200$	\\
Total gas mass				&	$3.2\times 10^{-2}\,$\msol		&	$2.3\times 10^{-2}\,$\msol		\\
Total dust mass				&	$8.1\times 10^{-4}\,$\msol		&	$1.1\times 10^{-4}\,$\msol		\\
$L_{\rm X}$					&	$7.94\times 10^{28}$~erg$\cdot$s$^{-1}$	& $1.4\times 10^{30}$~erg$\cdot$s$^{-1}$	\\
$T_{X}$ 						&	$7\times10^{7}$~K			& $3.2\times 10^{6}$~K	\\
$\zeta_{\rm cr}$				&	$5\times 10^{-17}$~s$^{-1}$	& $5\times 10^{-19}$~s$^{-1}$	\\
\hline
	&	\multicolumn{2}{c}{Derived abundances} \\
\hline
\chgas						&	$(0.1-1.5)\times 10^{-4}$		&	$(0.2-5.0)\times 10^{-6}$	 \\
\ohgas						&	$(0.14-1.4)\times 10^{-4}$		&	${\sim}1\times10^{-6}$		\\
\hline
\end{tabular}
\flushleft
\emph{Notes. }\gdrat\ is the gas to dust mass ratio. The stellar properties for HD~100546 are from the stellar abundance analysis in this work and differ somewhat from the literature values used in the disk modelling (Appendix~\ref{sec:hd546abuns} and Table~\ref{tab:uvlums}).
\end{table}

First, we fit the spectral energy distribution (SED), varying also $\delta_{\rm gas}$ and $\delta_{\rm dust}$ in the inner cavities of both disks. Combined, the SED, the resolved CO~$3$--$2$ emission profile, and the CO ladder and [\ion{O}{i}] lines constrain $\Sigma_{\rm dust}$, $h_{\rm c}$, $\psi$, $\gamma$, $f$, $\chi$, the properties of any large inner hole, and to some extent \gdrat. Fluxes or upper limits of HD lines constrain the mass of warm gas \citep[see also][]{Berginetal2013}. The spectral line profiles of CO and \ci\ further constrain the disk temperature and density structure, as well as the stellar mass and the disk inclination. Finally, modelling the CO ladder, the \ci, \cii\ and \oi\ lines constrains \chgas\ as well as the C/O ratio, although in HD~100546 the \cii\ and \oi\ lines are contaminated by envelope emission. We vary \ohgas\ either in lockstep with \chgas\ (fixed C/O ratio), or keep it fixed at $2.88\times 10^{-4}$ for variations of \chgas. The use of \texttt{DALI} for deriving \chgas\ is described in more detail in \citet{Kamaetal2016a}. The two main sources of uncertainty are the radial extent of the gas disk and the gas to dust ratio.

Both of our target disks show evidence for radial gaps in the millimetre opacity. We do not presently include these in our modelling, as the thermal structure of both disks is more strongly related to the small grains, traced by scattered light. Shallow radial dips and wave patterns are seen in scattered light \citep[e.g.][]{Ardilaetal2007, Debesetal2013}, but it is not clear that these represent large variations of UV heating and thus do not yet warrant inclusion in our modelling.

\begin{figure}[!ht]
\includegraphics[clip=, width=1.0\linewidth]{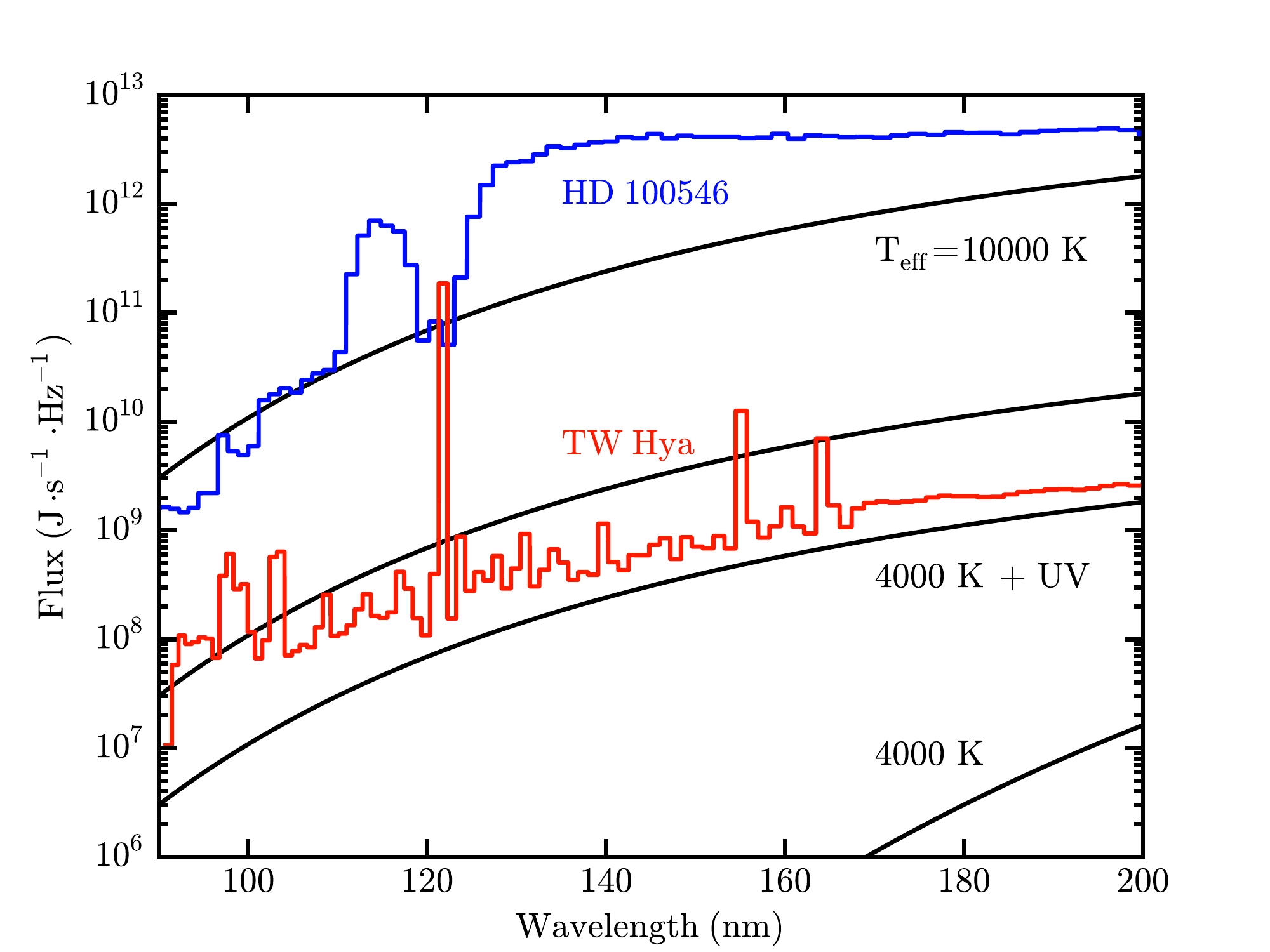}
\caption{The ultraviolet range of spectra used in this study for TW~Hya \citep[red curve,][]{Franceetal2014, Cleevesetal2015} and HD~100546 \citep[blue curve,][]{Brudereretal2012}, compared with blackbody photospheres for pre main sequence model stars with an age of $10\,$Myr \citep[black curves,][]{Tognellietal2011}. The two curves labelled $4000\,$K$+$UV show a $4000\,$K stellar photosphere with a $10000\,$K blackbody added for the accretion rates \mdot~$=10^{-9}$ and $10^{-8}\,$\msol$\,$yr$^{-1}$. See also Table~\ref{tab:uvlums}.}
\label{fig:spectra}
\end{figure}

Next, we describe the results of modelling each system. The range of well-fitting values for each parameter is listed in Table~\ref{tab:params}, and the disk structures and abundance maps are shown in Appendix~\ref{apx:linemaps}. The stellar spectra are from \citet{Brudereretal2012}, \citet{Franceetal2014} and \citet{Cleevesetal2015}, and their UV profile is shown in Fig.~\ref{fig:spectra}. Blackbody spectra of pre main sequence model stars from \citet{Tognellietal2011} for an age of $10\,$Myr are provided for comparison. To illustrate the importance of accretion power, we added UV excess to the $4000\,$K star corresponding to $10^{-9}$ and $10^{-8}\,$\msol$\,$yr$^{-1}$ of mass flux emitting its complete potential energy at the stellar surface at a temperature of $10000\,$K. The stellar luminosities are summarized in Table~\ref{tab:uvlums}. The range of well-fitting models was chosen by eye through consecutive parameter value refinements totalling about $100$ models per source.

\begin{table}[!ht]
\centering
\caption{The total and ultraviolet stellar luminosities for TW~Hya and HD~100546 and a set of model stars with blackbody spectra.}
\label{tab:uvlums}
\begin{tabular}{ l c c c }
\hline\hline
Spectrum		& CO ph.dissoc.	& Broadband UV			&	Total	\\
			& $91.2\ldots 110$~nm	&	$91.2\ldots 200$~nm	&	$L_{\rm tot}$	\\
			&	(\lsol)		&	(\lsol)	&	(\lsol)	\\
\hline
TW~Hya									& $2.8\times 10^{-4}$	&	$1.7\times 10^{-2}$	&	$0.28$	\\
HD~100546								& $1.5\times 10^{-2}$	&	$9.1\times 10^{\,0}$	&	$36$	\\
\hline
4000~K									& $8.6\times 10^{-12}$	&	$3.9\times 10^{-6}$	&	$0.25$	\\
$+$\mdot~$=10^{-8}\,$\msol$\,$yr$^{-1}$			& $1.8\times 10^{-4}$	&	$1.5\times 10^{-2}$	&	$0.48$	\\
$+$\mdot~$=10^{-9}\,$\msol$\,$yr$^{-1}$			& $1.8\times 10^{-5}$	&	$1.5\times 10^{-3}$	&	$0.28$	\\
10000~K									& $1.7\times 10^{-2}$	&	$1.5\times 10^{\,0}$	&	$23.0$	\\
\hline
\end{tabular}
\flushleft
\emph{Notes. }The luminosity of TW~Hya between $120$ and $123\,$nm, around the Lyman$\,\alpha$ line, is $1.2\times 10^{-2}\,$\lsol. The luminosity of HD~100546 derived from spectra at $\lambda{>}400\,$nm is $(25\pm7)\,$\lsol\ (Appendix~\ref{sec:hd546abuns}).
\end{table}

\subsection{HD~100546: carbon and oxygen are close to interstellar}\label{sec:hd100546}

\begin{figure*}[!ht]
\includegraphics[clip=, width=1.0\linewidth]{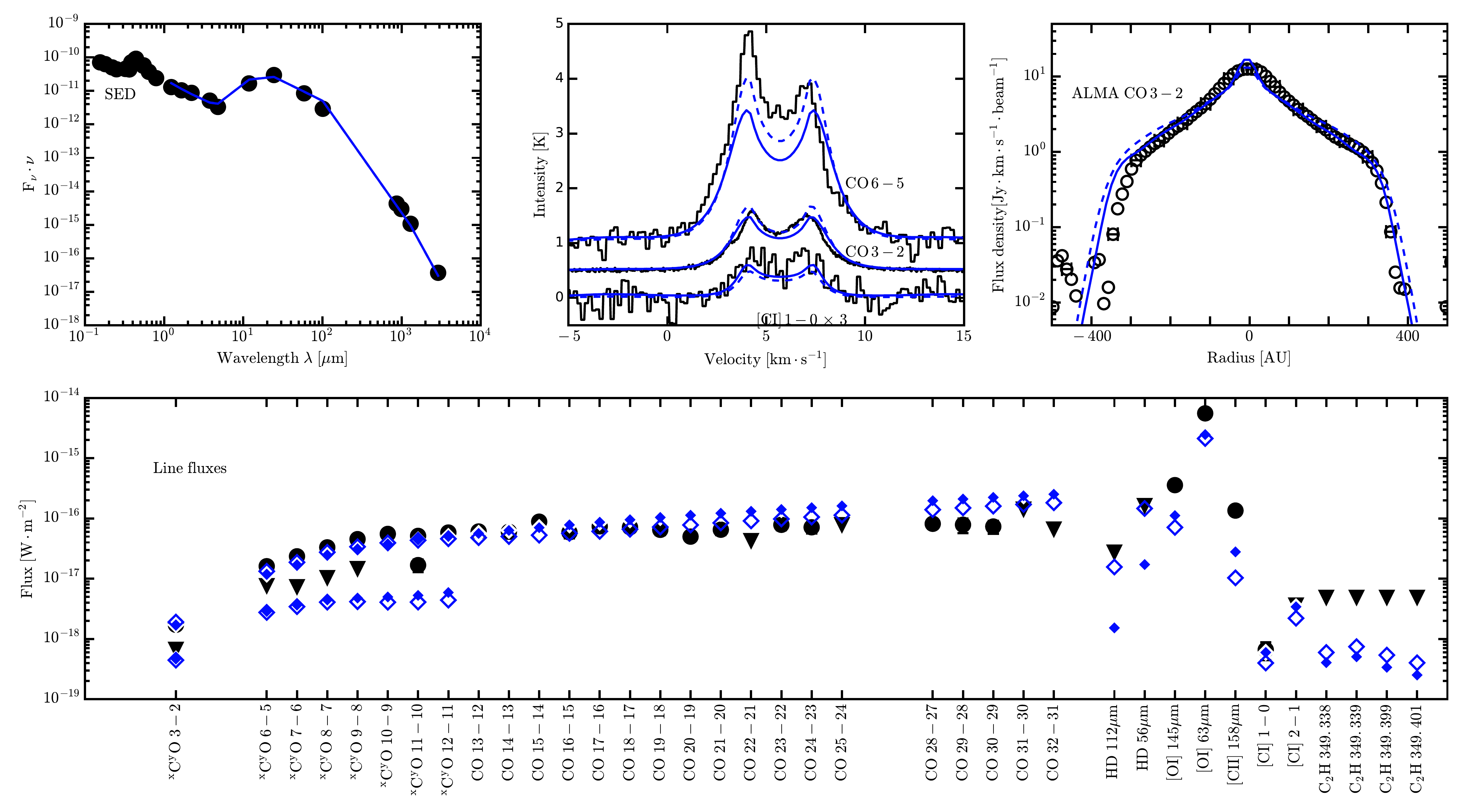}
\caption{HD~100546 observations and modelling. The panels show, from top left to bottom, the spectral energy distribution; spectrally resolved line profiles; spatially resolved integrated CO$\,3$--$2$ emission; and the full set of line fluxes used in the analysis. CO isotopolog lines for a given rotational transition are plotted at the same x-axis location. Two well-fitting models, with a gas to dust ratio \gdrat${=}10$ and \chgas${=}10^{-4}$ (solid blue) and \gdrat${=}100$ and \chgas${=}10^{-5}$ (dashed/open blue) are plotted on the observations (black) to illustrate the range of allowed parameters}. The data and model values for C$_{2}$H have been multiplied by a factor of $10^{3}$ for display purposes.
\label{fig:hd100546modelobs}
\end{figure*}

The results for HD~100546 are shown in Fig.~\ref{fig:hd100546modelobs}. Maps of the abundance and line emission of key species for the best-fit model with \chgas~$=10^{-4}$ and a \gdrat~$=10$ are shown in Fig.~\ref{fig:hd100546cbfs}. We show the spectral energy distribution (SED, top left panel), the \cionezero\ and CO~$3$--$2$ and $6$--$5$ line profiles (top middle), the radial cut of CO~$3$--$2$ intensity from ALMA (top right) and the compilation of spectral line fluxes (bottom panel). The disk is vertically extended and warm.

We obtain $10{\leq}$\gdrat${\leq}300$ for the outer disk. The upper limit relies on the HD~$1$--$0$ ($112\,\mu$m) and $2$--$1$ ($56\,\mu$m) upper limits calculated from the PACS spectrum of \citet{Fedeleetal2013b}. This is ${\lesssim}0.1\,$\msol\ or \gdrat~${\lesssim}300$, adopting our best fit dust mass of $8.1\times 10^{-4}\,$\msol. The corresponding surface density law is $\Sigma_{\rm gas}\lesssim150\,(r/50\,{\rm au})^{-1}\,$g$\,$cm$^{-2}$. Having \gdrat~${<}10$ requires \chgas~$\gtrsim 2\times 10^{-4}$ to match the CO lines, which however leads to the atomic carbon lines being overpredicted. 

We find \chgas$\,=(0.1-1.5)\times 10^{-4}$ and a C/O ratio that is solar to within a factor of a few. This uncertainty is mostly due to the limited constraints on the gas to dust ratio. An interstellar-like carbon abundance (\chgas$\sim10^{-4}$ in the diffuse ISM) provides a good fit if \gdrat${\sim}10$. For this model, additionally, the oxygen abundance must be $\gtrsim 10^{-4}$, otherwise the \ci\ line fluxes are substantially overpredicted as less carbon is bound into CO. Therefore, for the low gas to dust ratio and nominal \chgas\ solution, \ohgas\ must also be roughly interstellar. For \gdrat${=}100$, the best-fit value for \chgas\ is $\sim 2\times 10^{-5}$, assuming a solar C/O ratio. For this \gdrat, a carbon abundance of $10^{-4}$ overproduces the CO ladder by a factor of ten and the \ci\ detection by a factor of two. Decreasing the oxygen abundance such that the C/O ratio is above unity improves the fit on the CO ladder, but leads to an even more substantial overprediction of the \cionezero\ transition.

Using the dust disk model of \citet{Muldersetal2011} and a wide range of constraints on the disk and stellar properties, \citet{Brudereretal2012} modelled [\ion{C}{II}], [\ion{O}{I}] and CO ($J_{\rm u}\geq 14$) fluxes and [\ion{C}{I}] upper limits. They found gas-to-dust ratios from $20$ to $100$ and corresponding \chgas\ values of $1.2\times 10^{-4}$ (interstellar) to $1.2\times 10^{-5}$ (depleted from the gas by a factor of ten). Our results from above are consistent with these values.

Our analysis of the stellar photospheric abundance (Appendix~\ref{sec:hd546abuns}) is consistent with the star accreting carbon-rich gas, as would be expected in a system where volatile carbon is not depleted from the gas into planetesimals. We return to this in Section~\ref{sec:returnoftheice}.

The [\ion{O}{I}]$\,63\,\mu$m and $145\,\mu$m lines are always underproduced by factors of $2$ to $4$, and the spectrally resolved [\ion{C}{II}] emission by a factor of $\sim 5$. The latter line is single-peaked and very likely contains substantial emission from a residual circum-disk envelope \citep[e.g.][]{Fedeleetal2013a, Fedeleetal2013b, Dentetal2013}. As a consistency check, we calculate the envelope emission assuming \ngas$=10^{4}$~cm$^{-3}$, a spherical radius of $10^{3}\,$au and an abundance $10^{-4}$ of both O$^{0}$ and \cplus. Assuming \tkin$\,\sim50$ to $150$~K, the missing flux is recovered with such an envelope model.

We adopt the \gdrat${=}10$ and \chgas${=}10^{-4}$ model for HD~100546 in the rest of the paper, unless explicitly noted.

\begin{figure*}[!ht]
\includegraphics[clip=, width=1.0\linewidth]{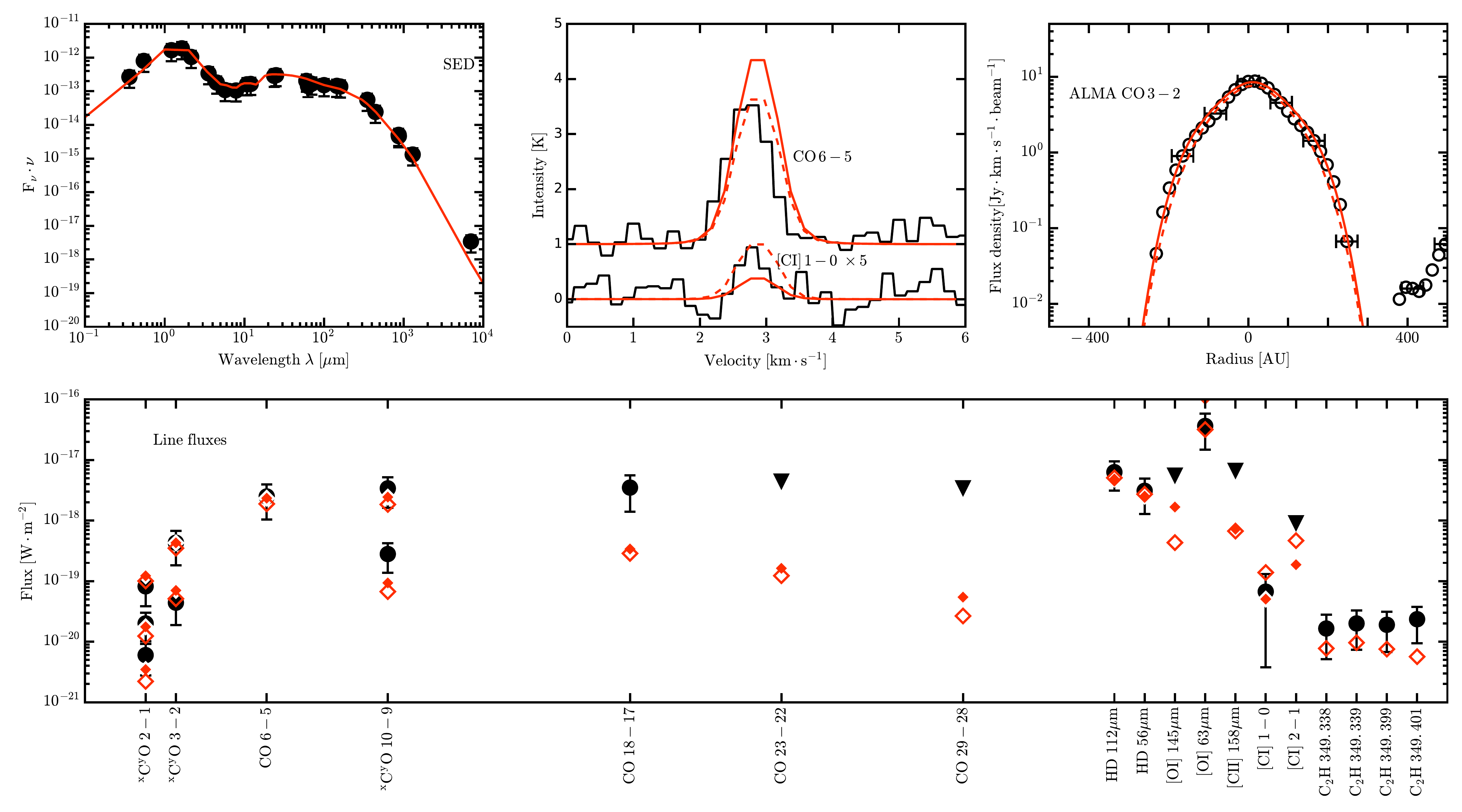}
\caption{TW~Hya observations and modelling. The panels show, from top left to bottom, the spectral energy distribution; spectrally resolved line profiles; spatially resolved integrated CO~$3$--$2$ emission; and the full set of line fluxes used in the analysis. CO isotopolog lines for a given rotational transition are plotted at the same x-axis location. Two well-fitting models, both with a gas to dust ratio \gdrat${=}100$ and \chgas${=}10^{-6}$ but with C/O${=}0.47$ (solid red) and $1.5$ (dashed/open red) are plotted on the observations (black) to illustrate the range of allowed parameters. For C/O${=}0.47$, the \oi$\,63\mu$m line model flux lies just above $10^{-16}\,$W$\,$m$^{-2}$ and the C$_{2}$H fluxes are a factor ${\approx}10^{3}$ below the observations.}
\label{fig:twhyamodelobs}
\end{figure*}

\subsection{TW~Hya: carbon and oxygen are underabundant}\label{sec:twhya}

The results for TW~Hya are shown in Fig.~\ref{fig:twhyamodelobs}. Maps of the abundance and line emission of key species for a typical well-fitting model, with \chgas~$=10^{-6}$, are shown in Fig.~\ref{fig:twhyacbfs}. We show the spectral energy distribution (SED, top left panel), the \cionezero\ and CO~$3$--$2$ and $6$--$5$ line profiles (top middle), the radial cut of CO~$3$--$2$ intensity from ALMA (top right) and the compilation of spectral line fluxes (bottom panel). Constrained by the spatially resolved CO~$3$--$2$ data \citep{Rosenfeldetal2012}, our best-fit models have a radially steeper temperature profile and smaller radius than the models of \citet{Cleevesetal2015}.

TW~Hya was the first disk where the gas mass was constrained using the HD~$112\,\mu$m transition \citep{Berginetal2013}. Our best-fit disk mass, $M_{\rm disk}=2.3\times 10^{-2}\,$\msol, is within a factor of two of the original range ($\geq5\times 10^{-2}\,$\msol). An additional feature of our model is that it reproduces the HD~$56\,\mu$m line, which will be presented in a forthcoming paper \citep{Fedeleetalinprep}.

The carbon and oxygen abundances are constrained to be low by the combination of \ci, \oi\ and low-$J$ CO lines and the HD~$112\,\mu$m line. The C$^{18}$O$\,2$--$1$ flux is another strong constraint on \chgas. However, our model does not yet include the isotopolog-selective photodissocation of CO, which may lead to an underestimation of \chgas\ or the total gas mass \citep{Miotelloetal2014b}. Since HD constrains the gas mass, in our case the uncertainty of a factor of a few falls on \chgas, but is somewhat mitigated by the complementary constraints from \ci\ and \cii. We will return to this in a companion paper (Miotello et al. in preparation). Overall, we find that both carbon and oxygen must be substantially underabundant in the gas, with a \chgas\ ratio of $(0.2-5.0)\times 10^{-6}$. The \ohgas\ ratio is constrained to be ${\sim}10^{-6}$. A higher value strongly overpredicts the [\ion{O}{i}]~$63\,\mu$m flux, while a much smaller value leads to an excess of \catom\ and C$_{2}$H, as carbon normally bound into CO becomes available.

The \emph{Herschel}/PACS upper limits on the high-$J$ CO lines \citep{Kampetal2013} also constrain \chgas~$<10^{-4}$. The CO~$23$--$22$ line is formally detected, but a recent re-analysis (D. Fedele, priv.comm.) suggests it is blended with an H$_{2}$O line and we thus treat it as an upper limit.

Based on C$^{18}$O emission and the HD gas mass, an elemental gas-phase carbon and oxygen deficiency of a factor of $10$--$100$ was already inferred for this disk \citep{Hogerheijdeetal2011, Berginetal2013, Favreetal2013, Duetal2015}. Our models confirm these results.

\subsection{Modelling of C$_{2}$H and constraining the C/O ratio}\label{sec:c2h}

C$_{2}$H is another tracer of the carbon chemistry in the disk surface layers. The hydrocarbon chemistry in our network extends only up to C$_{2}$H and C$_{2}$H$_{2}$. In addition to data from UMIST06, we include the reactions $\rm C_{2}H+H_{2}\rightarrow C_{2}H_{2}+H$, with an activation barrier of E$\rm_{a}=624\,$K (UMIST12), and $\rm C_{2}+H_{2}\rightarrow C_{2}H+H$, with E$\rm_{a}=1469\,$K \citep{Cernicharo2004}, but their effect on the results is not substantial. We note the latter reaction has been studied at high pressure, and the rate may not apply at disk conditions. The ring-like morphology of C$_{2}$H emission around TW~Hya led \citet{Kastneretal2015} to invoke the photodestruction of hydrocarbons and small carbonaceous grains as a source. We have not included this process in our network.

In our best-fit models, we keep C/O ratio at the solar value. This underproduces the C$_{2}$H fluxes in TW~Hya by a factor of $100$, while the upper limits for C$_{2}$H in HD~100546 are a factor $1000$ above the models. The C$_{2}$H abundance may be enhanced or decreased depending on the carbon-to-oxygen ratio of the gas. If C/O~$\gtrsim1$, hydrocarbons start to become much more abundant. To test this, we ran models with a decreased \ohgas, making C/O~$=1.5$. For TW~Hya, this makes the modelled C$_{2}$H fluxes consistent with the observations and somewhat improves the match for \oi. We plot the results in Fig.~\ref{fig:twhyamodelobs}. For HD~100546, however, the models now lie a factor of fifty above the observational upper limits. Qualitatively, we can conclude that the models suggest a stronger depletion of oxygen than carbon in the outer disk gas of TW~Hya, while for HD~100546 there is no evidence for such a differential depletion. Full modelling of C$_{2}$H is outside the scope of this paper.

\begin{figure}[!ht]
\includegraphics[clip=,width=1.0\columnwidth]{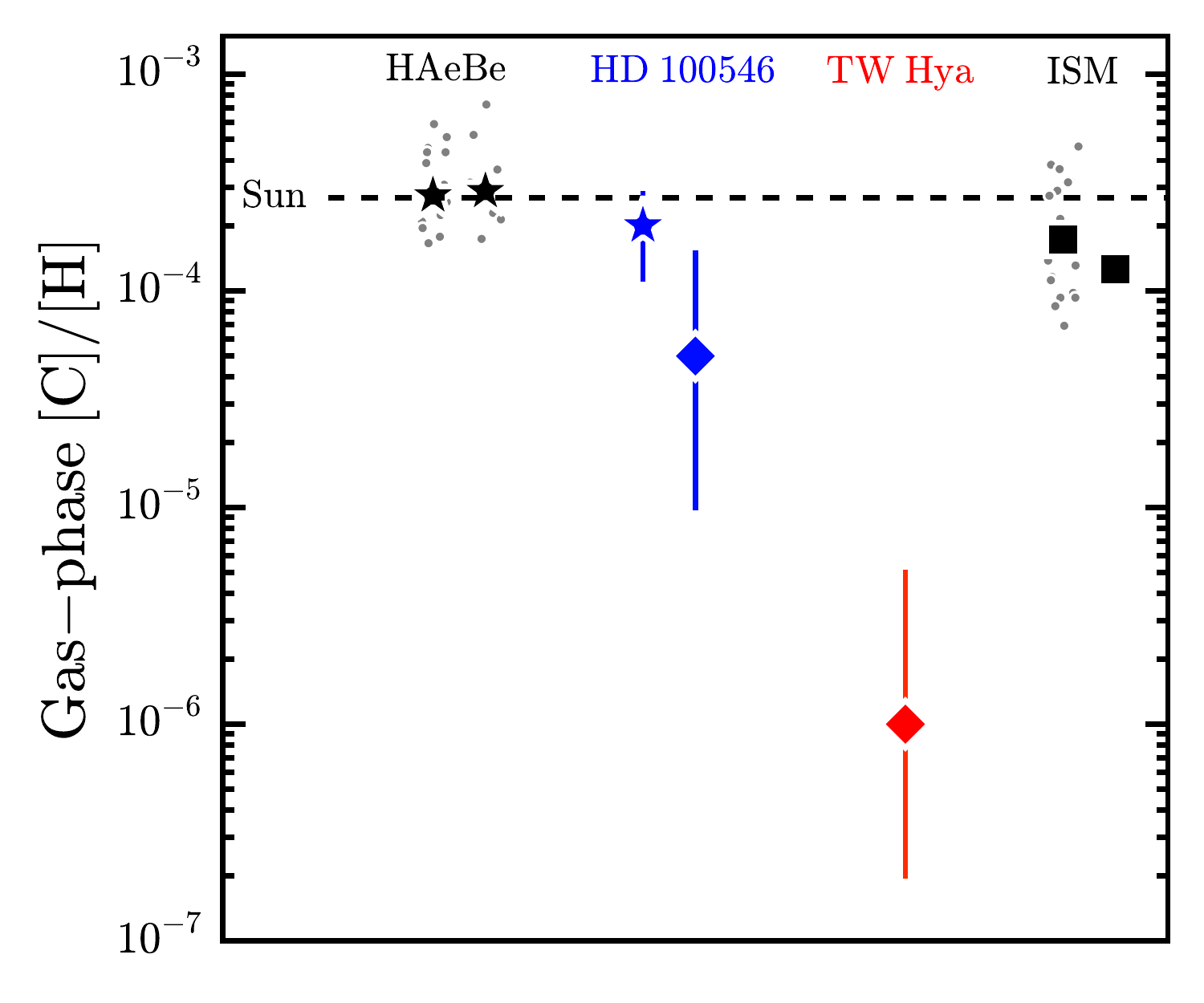}
\caption{The gas-phase carbon abundance, compared with that of the Sun \citep[dashed line,][]{Asplundetal2009}. From left to right: photospheric values for Herbig~Ae/Be stars \citep[stars]{AckeWaelkens2004, Folsometal2012}; the stellar and disk values for HD~100546 (star and diamond, this work); the disk of TW~Hya (diamond, this work); and lines of sight through the diffuse interstellar medium \citep{Cardellietal1996, Parvathietal2012}. For the Herbig~Ae/Be and ISM samples, the median (black symbols) and the full dataset (small gray circles) are shown.}
\label{fig:abunC}
\end{figure}

\section{The locking and release of volatiles in disks}\label{sec:discussion}

We have confirmed earlier findings by \citet{Brudereretal2012}, \citet{Favreetal2013}, and \citet{Duetal2015} that the gas-phase volatile carbon and oxygen are underabundant by two orders of magnitude in the TW~Hya disk, while they are roughly interstellar or moderately underabundant in HD~100546. In Fig.~\ref{fig:abunC}, we compare the derived \chgas\ ranges for these disks with the photospheric abundance of carbon in HD~100546 and in a large sample of other Herbig~Ae/Be stars \citep{AckeWaelkens2004, Folsometal2012}, as well as with the interstellar and solar carbon abundances \citep{Asplundetal2009, Cardellietal1996, Parvathietal2012}. The photospheric value for HD~100546 is consistent with the rest of the Herbig star sample, while the \chgas\ ratio in the disk gas is consistent with interstellar or slightly sub-interstellar values. We find no solution for the TW~Hya disk in which the \chgas\ value is close to interstellar.

\begin{figure}[!ht]
\includegraphics[clip=,width=1.0\columnwidth]{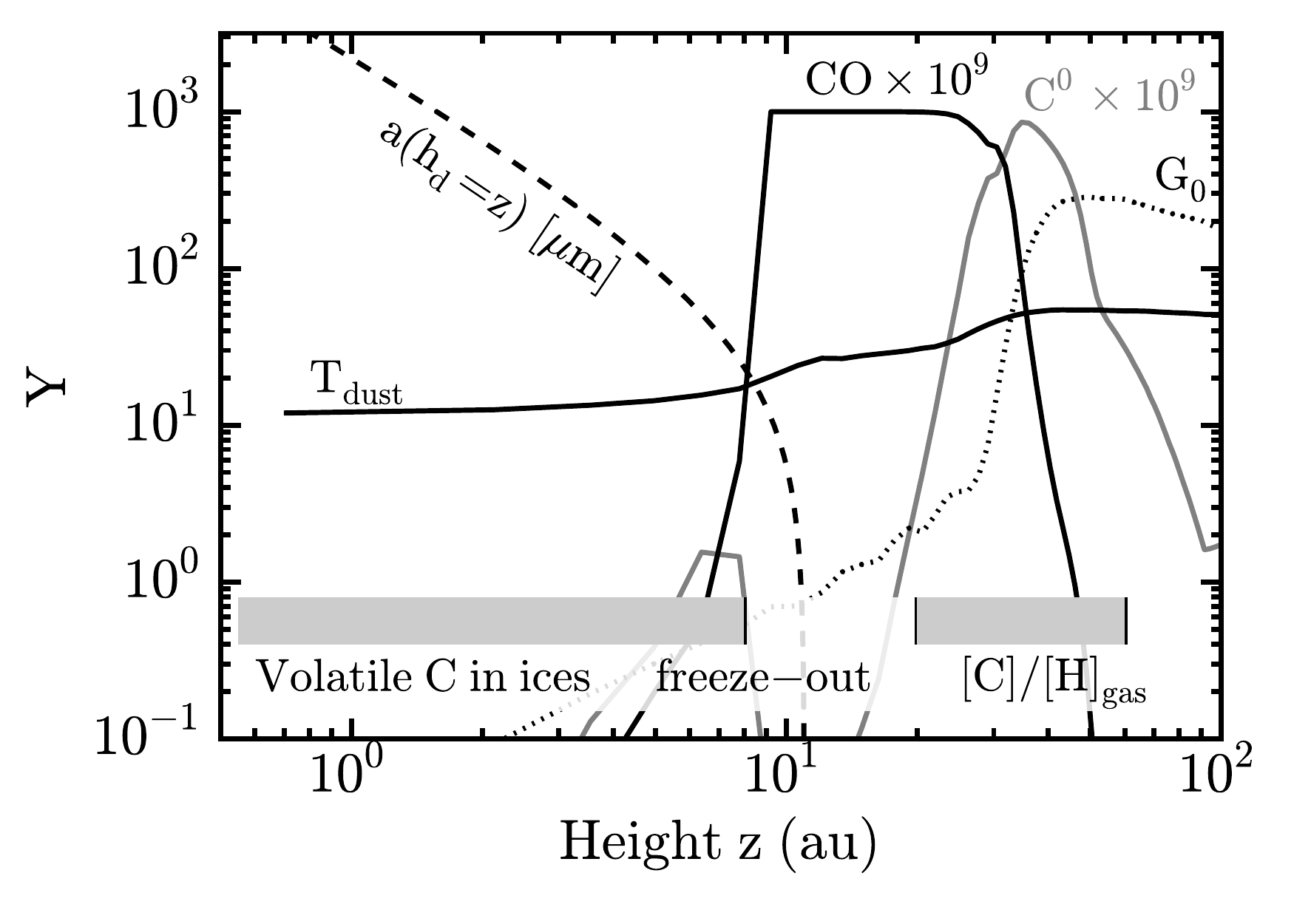}
\caption{The vertical structure of the TW~Hya model at $r=100\,$au. $\rm Y$ denotes either the dust temperature ($T_{\rm dust}\,[{\rm K}]$), UV-field ($G_{0}$), gas-phase CO and \catom\ abundance, or the grain size ($a\,[\mu$m$]$) of particles with a scaleheight $h_{\rm d}$ equal to a given height $z$ (see Eq.~\ref{eq:hdz}). Vertical mixing can transport carbon atoms from the \catom\ layer to the freeze-out zone.}
\label{fig:zcut}
\end{figure}

The identified pattern -- a strong underabundance in a cold disk and at most moderate underabundance in a warm disk -- is consistent with a scenario where vertical mixing transports gas from the disk surface into the cold midplane, where CO, \htwoo, hydrocarbons and perhaps other C and O carriers can freeze out onto dust grains. We show the main aspects of the depletion for TW~Hya in Fig.~\ref{fig:zcut}, where the relevant parameters are shown as a function of height at $r{=}100\,$au. Our measurement of \chgas\ relies strongly on \ci\ and CO emission, and probes mainly the outer disk atmosphere. To remove carbon atoms from this region, they must be transported to the cold midplane, where some of them must then remain. We propose that the systematic loss of carbon is due to freeze-out onto grains which are sufficiently decoupled to have scaleheights smaller than the CO freeze-out height (CO snow-surface, the vertical location where \tdust~$=17\,$K). Below, we model this sequestration process and constrain the nature of the main carrier species that freeze out. Finally, we discuss evidence for the evaporation of volatile ices in the warm inner disk of TW~Hya.

\begin{figure}[!ht]
\includegraphics[clip=,width=1.0\columnwidth]{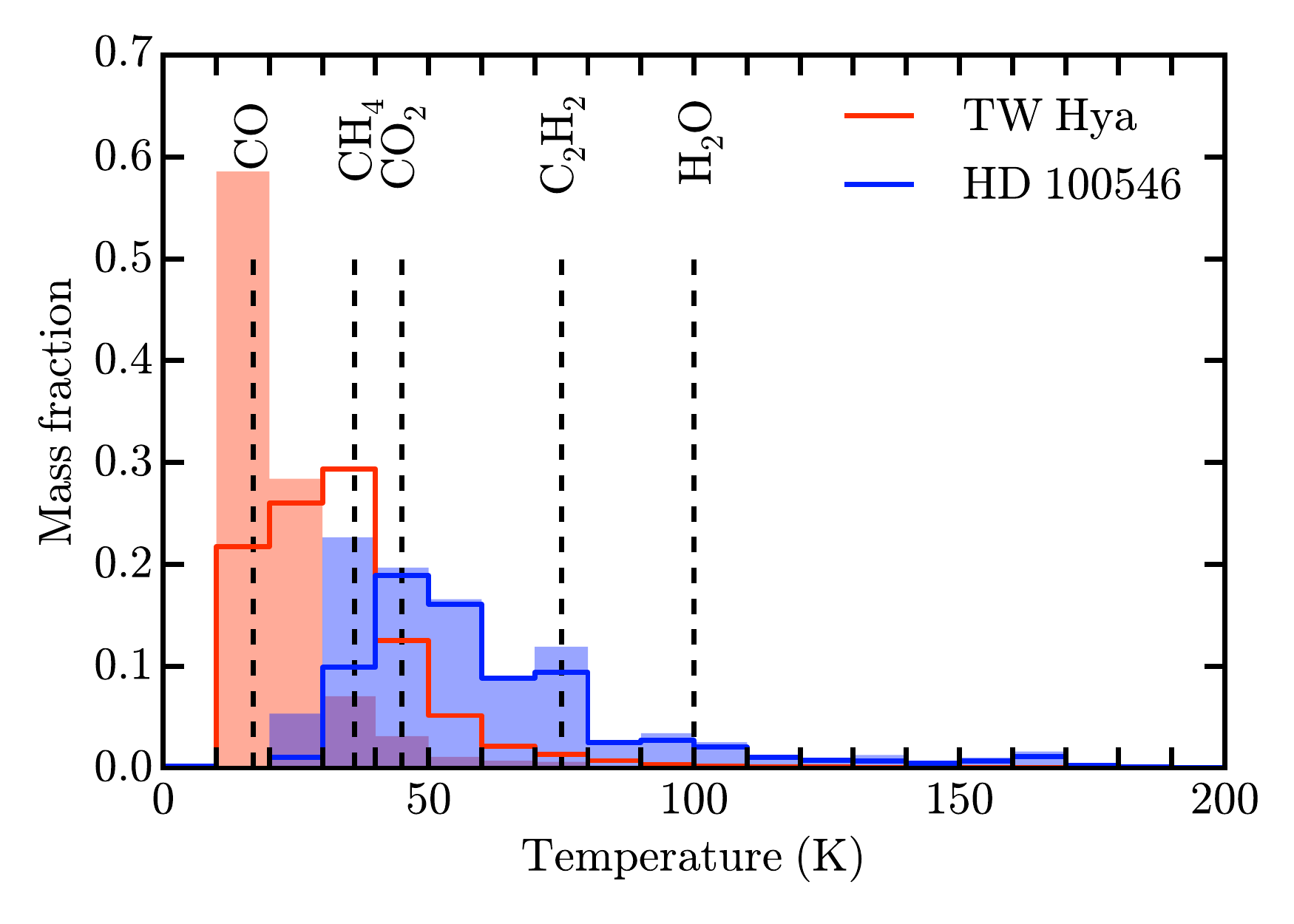}
\caption{Histograms of the fractional gas (solid lines) and dust mass (shaded) in $10\,$K temperature bins for the full disk of TW~Hya (red) and HD~100546 (blue). Freeze-out temperatures of the main species for C and O loss from the gas to ices are marked with vertical dashed lines.}
\label{fig:masshisto}
\end{figure}

\subsection{In what form does volatile carbon freeze out?}\label{sec:origindiscussion}

What species are the main gas-phase carriers that contribute to the loss of carbon to midplane ices? One possible explanation is that the CO$+$He$^{+}\rightarrow {\rm C}^{+}$ reaction can liberate carbon from CO in the disk atmosphere. Via reactions with H$_{2}$, electrons, and \catom, hydrocarbon species are then produced \citep{Aikawaetal1999}. These may become important reservoirs of gas-phase carbon \citep{Berginetal2014, FuruyaAikawa2014}. Alternatively, the sequestration into ices may start with the mixing and freezeout of CO itself. In Fig.~\ref{fig:masshisto}, we show histograms of the dust and gas mass fraction in temperature bins of $10\,$K for our best models of both disks. We can now test which species dominate the loss of volatile carbon from the gas phase in the outer disk. The main agent of underabundance should condense in the colder disk (TW~Hya), but not in the warmer one (HD~100546). Note that the freezeout is controlled by the dust temperature.

Using C$_{2}$H$_{2}$ as a lower limit for large hydrocarbons, for which the sublimation temperature generally increases with size \citep[e.g.][]{Chickos1986}, we see that this class is ruled out as a major gas-phase C carrier. C$_{2}$H$_{2}$ is easily able to freeze out in both disks. Some fraction of carbon may be lost as CH$_{4}$ and CO$_{2}$, but they are likely not the dominant gas-phase carriers, as both can freeze out in the outer disk of HD~100546. Most gas-phase carbon atoms in the HD~100546 disk have, therefore, not passed through a complex hydrocarbon molecule during the 10~Myr life of the system. This is consistent with chemical models, where CH$_{4}$ production usually requires long timescales and CO$_{2}$ is produced by thermal or photo-processing of CO-, H$_{2}$O-, and OH-rich ices \citep[e.g.][]{Willacyetal1998, Furuyaetal2013, Walshetal2014a}. Observable gas-phase complex organic molecules in the protostellar source Orion~KL only contain $\lesssim15$\% of the organic carbon locked in Solar System meteorites and comets, suggesting that large amounts of complex organics are produced from simple carbon-bearing species during the protoplanetary disk stage \citep{Berginetal2014}. All evidence thus indicates that carbon is lost from the outer disk gas predominantly as CO, which is able to freeze out in a large mass fraction of the TW~Hya disk, but not in HD~100546. Further grain surface processing of carbon in ices starts from this parent species. The formation of large amounts of complex organics from this is only possible in disks cold enough for CO freeze-out.

Figure~\ref{fig:masshisto} raises the question of the gas-phase oxygen abundance in the HD~100546 disk, as both disks easily allow for near-complete H$_{2}$O freeze-out. This should facilitate the sequestering of oxygen into midplane ices. In TW~Hya, most water ice was indeed inferred to be trapped in the midplane, based on the low H$_{2}$O line fluxes observed with \emph{Herschel} \citep{Hogerheijdeetal2011}. Upper limits on H$_{2}$O emission towards HD~100546 \citep{Sturmetal2010, Thietal2011b, Meeusetal2012, Fedeleetal2013b} still allow an interstellar gas-phase oxygen abundance, \ohgas~$=2.88\times 10^{-4}$, given our best-fitting disk model. A recent H$_{2}$O detection in HD~100546 will provide stronger constraints on \ohgas\ \citep[Hogerheijde et al. in preparation, see also][]{vanDishoecketal2014}.

\subsection{An analytical model of volatile locking}\label{sec:lockmodel}

Two orders of magnitude of gas-phase carbon have been lost from the gas disk of TW~Hya, and the processes responsible are likely general and operating on large scales. We present an analytical model for the loss of volatile carbon in a disk, relating a progressing depletion to vertical mixing, freeze-out (ostensibly of CO), and the dust size distribution.

Let $r$ be a radial location in the disk. We assume that gas and the small grains that couple to it at $r$ are constantly vertically mixed, such that they cycle between the disk atmosphere and the midplane. We can remain agnostic as regards the vertical mixing mechanism, as long as the round-trip mixing time $t_{\rm mix}$ can be estimated. In each mixing cycle, let a fraction $\Delta_{\rm X}$ of volatiles get permanently locked in the midplane. Assuming the mixing or diffusion cycle lasts $t_{\rm mix}\,$years and the process repeats for a duration of $t_{\rm tot}\,$years, we have $N=t_{\rm tot}/t_{\rm mix}$ cycles. The initial abundance $X_{\rm 0}$ of an element in the disk surface gas will then be depleted to a lower abundance
\begin{equation}\label{eq:x1x0}
X_{\rm 1}=X_{\rm 0}\times (1-\Delta_{\rm X})^{N}.
\end{equation}

\begin{figure*}[!ht]
\centering
\includegraphics[clip=,width=1.0\linewidth]{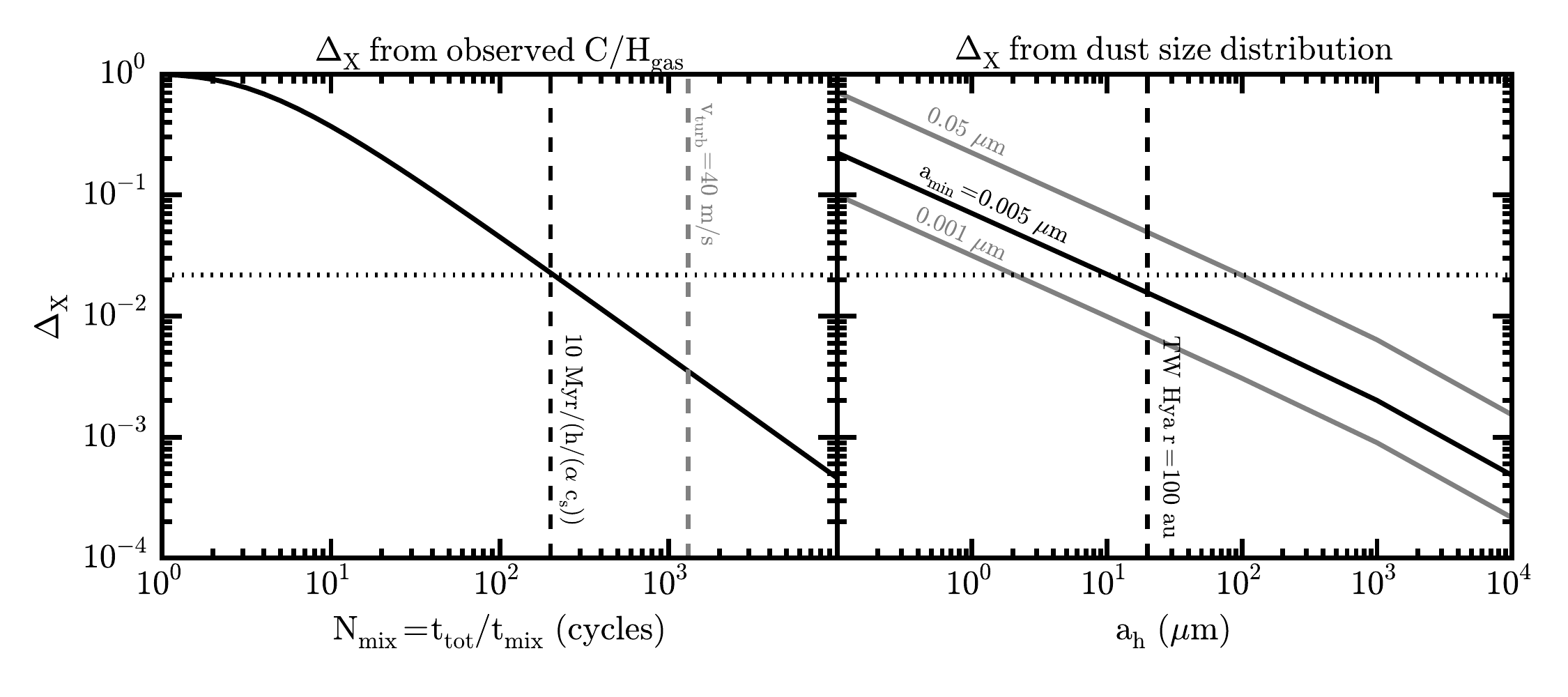}
\caption{A model of volatile loss from the disk atmosphere.
\emph{Left-hand panel: }A fraction $\Delta_{\rm X}$ of volatile atoms are lost in the midplane during each of $N_{\rm mix}$ mixing cycles. Eq.~\ref{eq:x1x0} is shown for a depletion $X_{0}/X_{1}{=}100$ (solid curve). A theoretical turbulent mixing timescale gives $N_{\rm mix}{=}200$ (black dashed line) and $\Delta_{\rm X}{=}2\times10^{-2}$ (dotted line), consistent with an upper limit of $v_{\rm turb}{\leq}40\,$m$\,$s$^{-1}$ on the speed of turbulent mixing in TW~Hya, which gives $N_{\rm mix}{\leq}1300$ \citep[gray dashed line,][]{Hughesetal2011}. \emph{Right-hand panel: }$\Delta_{\rm X}$ from Eq.~\ref{eq:Dxdust}, taking $a_{\rm max}{=}10\,$cm and $a_{\rm min}{=}0.001$, $0.005$, or $0.05\,\mu$m (solid lines). The smallest grain size settled below the CO freeze-out height at $r{=}100\,$au in our TW~Hya model is also shown (dashed line, Eq.~\ref{eq:hdz}).}
\label{fig:deltaX}
\end{figure*}

The age of TW~Hya is $t_{\rm tot}=10\,$Myr. We adopt a turbulent mixing time $t_{\rm mix}=h^{2}/\nu_{\alpha}$, where $\nu_{\alpha}=\alpha\,c_{\rm s}\,h$ is the parameterized viscosity of \citet{ShakuraSunyaev1973}, and further set $\alpha=0.01$. Taking the disk atmosphere height $h$, defined here as the region with peak \catom\ abundance, at $r=100\,$au to be $30\,$au (Fig.~\ref{fig:twhyacbfs}) and the sound speed $c_{\rm s}=300\,$m$\,$s$^{-1}$, we find $t_{\rm mix}=5\times 10^{4}\,$yr. The left-hand panel of Fig.~\ref{fig:deltaX} shows $\Delta_{\rm X}$ and $N$ which satisfy $X_{0}/X_{1}=100$ (Eq.~\ref{eq:x1x0}). An underabundance of a factor of $X_{\rm 0}/X_{\rm 1}=100$ over $10\,$Myr corresponds to $N_{\rm mix}=200$ and a fractional loss per mixing cycle of $\Delta_{\rm X}=2\times 10^{-2}$. In one cycle, $98\,$\% of the atoms either do not freeze out, or freeze onto grains that are well coupled to the gas and return to the warm disk atmosphere where they desorb.

We interpret the carbon loss term, $\Delta_{\rm X}$, as freezeout onto large grains which are settled below the height where volatile carbon freezes out. The likeliest candidate carrier is CO, which freezes out in TW~Hya at $17\,$K \citep{Qietal2013}. Volatiles that freeze onto grains which have a scaleheight equal to or smaller than the CO snow-surface, where $h_{\rm d}=z_{\rm freeze-out}$ (this happens at $T_{\rm dust}=17\,$K; see also Eq.~\ref{eq:hdz} and Fig.~\ref{fig:zcut}), stay locked as ices in the midplane, while ices formed on smaller grains eventually get mixed back to the warm surface layers where they desorb. The loss term $\Delta_{\rm X}$ is equal to the fraction of dust surface area contributed by grains which are settled below the CO snowline. We adopt a standard dust size distribution of the form $dn/da=C\times a^{-3.5}$ between $a_{\rm min}=0.005\,\mu$m and $a_{\rm max}=10\,$cm, where $C$ is a constant, and denote the size of the smallest grains which satisfy $h_{\rm d}(a_{\rm h})=z_{\rm freeze-out}$ as $a_{\rm min}\leq a_{\rm h}\leq a_{\rm max}$. Note that the maximum grain size here exceeds that used in the opacities in Section~\ref{sec:twhya}. Assuming that the total surface area of dust above the CO snowline is negligible compared to that below it, we arrive at

\begin{equation}\label{eq:Dxdust}
\Delta_{\rm X} = \left( 1 + \frac{1 - \left(a_{\rm h}/a_{\rm min}\right)^{0.5}}{\left(a_{\rm h}/a_{\rm max}\right)^{0.5} - 1}\right)^{-1}
\end{equation}

This expression for $\Delta_{\rm X}$ is plotted in the right-hand panel of Fig.~\ref{fig:deltaX} as $\Delta_{\rm X}(a_{\rm h})$. Based on \citet{YoudinLithwick2007}, we also plot the grain size at which the dust scaleheight equals the CO snow-surface of $z{\approx}8\,$au. This is $a_{\rm h}=20\,\mu$m, obtained from

\begin{equation}\label{eq:hdz}
\frac{h_{\rm d}(a_{\rm h})}{h_{\rm gas}} = \left[ \left( 1 + \frac{ \Omega\, t_{\rm fric}(a_{\rm h}) }{ \alpha }\right)\, \left( \frac{ 1 + 2\,\Omega\, t_{\rm fric}(a_{\rm h}) }{ 1 + \Omega\, t_{\rm fric}(a_{\rm h}) } \right) \right]^{-1/2},
\end{equation}

where $h_{\rm gas}=11\,$au is the gas scaleheight at $100\,$au in our TW~Hya model, and $\Omega$ is the Keplerian frequency at the radial location, $\Omega(r{=}100\,{\rm au})\approx (10^{-10}/\pi)\,$s$^{-1}$ in our case. The stopping time in the Epstein regime is
\begin{equation}\label{eq:epsteinstop}
t_{\rm fric}(a) = \frac{\rho_{\rm d}}{\rho_{\rm gas}}\,\frac{a}{c_{\rm s}},
\end{equation}
where $\rho_{\rm d}$ is the dust material density, and $\rho_{\rm gas}$ is the gas density. We adopt $\rho_{\rm dust}{=}3\,$g$\,$cm$^{-3}$ and, from Fig.~\ref{fig:twhyacbfs}, we get $\rho_{\rm gas}{\approx}10^{-15}\,$g$\,$cm$^{-3}$ at $r{=}100\,$au.

Figure~\ref{fig:deltaX} shows that the observationally determined $\Delta_{\rm X}{=}2\times 10^{-2}$ is consistent with that predicted from a settled dust population whose vertically integrated size distribution is $dn/da=C\cdot a^{-3.5}$, with a given $a_{\rm min}$. However, $\Delta_{\rm X}$ is currently likely much smaller, as we discuss next. While mm-to-cm grains have been directly detected in the inner disk of TW~Hya, the disk is strongly depleted of millimetre-sized grains outside of $60\,$au \citep{Wilneretal2005, Andrewsetal2012}. Thus, volatile loss onto large grains is likely not very efficient at $r{=}100\,$au at present, which implies that carbon was removed more efficiently in the past, i.e. $N_{\rm mix}<200$ and $\Delta_{\rm X}>2\times10^{-2}$. Our results are consistent with $a_{\rm min}\gtrsim 0.005\,\mu$m.

If the carbon loss happened on the same timescale as the loss of $1\,$cm grains due to radial migration -- $t_{\rm tot}{\sim} 10^{5}\,$years -- we find $\Delta_{\rm X}=0.9$. The timescale for coagulation to $1\,$cm size  -- $t_{\rm tot}{\sim}10^{6\,}$years -- yields $\Delta_{\rm X}=0.21$. The dust growth timescale may be substantially shorter than this \citep[e.g.][]{DullemondDominik2005}, in which case $\Delta_{\rm X}>0.21$ in the past. 

The assumptions and predictions of our simple volatile loss model need to be tested with measurements of \chgas\ in disks covering a range of ages and average temperatures, and we are pursuing such a follow-up programme. It is possible that grain growth provides an additional efficiency boost to the volatile loss process, by trapping some of the ices formed on small, coupled grains in larger particles via coagulation before the small grains can be mixed back to the disk surface \citep{Duetal2015}. The timescales of dust growth and vertical mixing are similar, within an order of magnitude, and their competition is not yet well observationally constrained. Direct measurements of the level of turbulence with ALMA show promise for remedying this \citep[e.g.][]{Hughesetal2011, Guilloteauetal2012, Flahertyetal2015}.

\subsection{The return of volatiles to the gas in the inner disk}\label{sec:returnoftheice}

Even around TW~Hya, large ice-covered grains would eventually radially migrate close enough to the star for the volatiles to evaporate back to the gas phase. The current data do not allow to robustly measure gas-phase elemental abundances in the inner disk for our targets to directly check this hypothesis. We propose to constrain the inner disk gas composition by studying the gas accreting onto the central stars.

Early-type stars, including Herbig~Ae/Be stars, have radiative envelopes with mixing timescales that are longer than the timescale for accreting the mass of the stellar photosphere, and their observed photospheric abundances reflect the composition of the accreted material \citep{TurcotteCharbonneau1993, Turcotte2002}. The photospheric abundances of Herbig~Ae/Be stars were recently shown to be a predictor of the inner disk gas and dust content \citep{Kamaetal2015b}. This is in contrast to T~Tauri stars, which have convective envelopes, where mixing rapidly erases accretion signatures. 

Using tested stellar abundance analysis methods for Herbig stars \citep[e.g.][]{Folsometal2012}, we have derived the stellar abundances for HD~100546 (see Appendix~\ref{sec:hd546abuns} for a full description). HD~100546 displays the $\lambda$~Bo\"{o}tis phenomenon, meaning it has approximately solar C, N and O abundances but is strongly depleted in refractory elements such as Fe, Si, Mg and Al. The $\lambda$~Bo\"{o} phenomenon is thought to be due to the preferential accretion of gas rather than dust \citep{VennLambert1990}. Intriguingly, \citet{Gradyetal1997} inferred a relative enhancement of the C/Fe ratio in gas clouds accreting onto HD~100546. They further found siderophile elements overall to be depleted in the accreting gas. The above results are consistent with most C being in the gas phase in this disk, and thus being readily accreted onto the star. It also confirms that abundance ratios in the accretion flows do trace the composition of the accreting material.

Photospheric abundances for T~Tauri stars are less sensitive to the composition of the currently accreting material, as noted above. Thus the stellar abundances for TW~Hya would not reflect the inner disk composition. The connection between photospheric abundances and those in the accreting gas around HD~100546 suggests a way around this limitation, if the composition of the gas accreting onto TW~Hya can be determined.

Studies of the UV spectrum of gas accreting onto TW~Hya have found the [\ion{C}{iv}]/[\ion{Si}{iv}] ratio to be about an order of magnitude higher than in most other T~Tauri stars, possibly owing to the accretion of silicon-poor (carbon-rich) material \citep[e.g.][]{Ardilaetal2002a, Herczegetal2002, Ardilaetal2013}. This appears inconsistent with \chgas~$=10^{-6}$ and points to the sublimation of carbonaceous species in the inner disk. In order for the refractory silicates to not be accreted, they must remain trapped between the dust sublimation radius (close to the stellar surface for TW~Hya) and the radius at which the carbon-rich ices evaporate (${\lesssim}30\,$au for CO, ${\lesssim}4\,$au for CH$_{4}$). This implies an ongoing build-up of volatile-poor large dust particles or planetesimals within $30\,$au at most.

C$_{2}$H$_{2}$ and H$_{2}$O were not detected in \emph{Spitzer}/IRS mid-infrared spectroscopy of TW~Hya by \citet{Najitaetal2010}. These observations probe the inner few astronomical units, much further out than the few stellar radii probed by UV observations of accreting gas. The mentioned species are commonly seen in such spectra and may originate from sublimation of ices or from gas-phase production mechanisms \citep[e.g.][]{Pascuccietal2009, Pascuccietal2013, Walshetal2015}. Their non-detection may indicate that the dust as well as the gas are depleted in the inner disk, and does not necessarily mean that the volatiles have not evaporated at all. Such inner disk depletions, with dust orders of magnitude more deficient than gas, have been found in ALMA studies of several transitional disks around Herbig Ae/Be stars \citep[e.g.][]{Brudereretal2014, Zhangetal2014, vanderMareletal2015}, and no doubt high spatial resolution observations of TW~Hya will soon shed more light on the matter.

\section{Conclusions}\label{sec:conclusions}

Based on new detections of atomic carbon emission from the disks around HD~100546 and TW~Hya, we have revisited the gas-phase carbon abundance in the atmospheres of these disks. The analysis makes use of the DALI code and a range of continuum and spectral line data from the literature. For HD~100546, we also studied the stellar photospheric abundances, and for both systems employed various UV/NIR constraints to discuss elemental abundances in the inner disk.

\begin{enumerate}

\item{We present the first unambiguous detections of atomic carbon emission from protoplanetary disks, in the sources HD~100546 and TW~Hya.}

\item{For both disks, we fitted fully parametric physical models to the spectral energy distribution, spatially resolved CO emission, and a range of far-infrared emission lines. The most important difference from past physical models is that spatially resolved CO line emission is also reproduced.}

\item{We confirm earlier findings that the HD~100546 disk atmosphere has either an interstellar gas-phase carbon abundance (\chgas${\approx}10^{-4}$, if \gdrat${=}10$) or is moderately depleted (\chgas${\approx}10^{-5}$, if \gdrat${=}100$); and that the TW~Hya disk atmosphere is two orders of magnitude deficient in gas-phase volatiles (\chgas${\approx}10^{-6}$, \ohgas${\approx}10^{-6}$, and C/O${>}1$).}

\item{Upper limits on HD emission for HD~100546 put an upper limit of $300$ on the gas to dust ratio in the disk.}

\item{The star HD~100546 has solar-like photospheric abundances of volatile elements (C, N, O) but a strong depletion of rock-forming elements (e.g. Fe, Mg, Si).}

\item{Most carbon atoms in the HD~100546 disk gas have not passed through a hydrocarbon more complex than CH$_{4}$ during the lifetime of the system.}

\item{An underabundance of volatile elements in the disk atmosphere can develop when gas is mixed through the midplane and the temperature there allows freezeout of major carrier species onto settled grains. An analytical formulation of this process is presented in Section~\ref{sec:lockmodel}.}

\item{The volatiles return to the gas in the inner disk, as evidenced by their enhanced abundance relative to refractory elements in the gas accreting onto both stars.}

\end{enumerate}

\begin{acknowledgements}
We thank Catherine Walsh for sharing her ALMA data on HD~100546; Steve Doty and Sebastiaan Krijt for stimulating discussions; Sean Andrews and Gijs Mulders for sharing their compiled spectral energy distributions; and the ALLEGRO ALMA node in Leiden for their help. This work is supported by a Royal Netherlands Academy of Arts and Sciences (KNAW) professor prize, by the Netherlands Research School for Astronomy (NOVA), and by the European Union A-ERC grant 291141 CHEMPLAN. CPF acknowledges support from the French ANR grant ``Toupies: Towards understanding the spin evolution of stars''. DF acknowledges support from the Italian Ministry of Education, Universities and Research project SIR (RBSI14ZRHR). This publication is based on data acquired with the Atacama Pathfinder Experiment (APEX), proposals 093.C-0926 and 093.F-0015, and on observations made with ESO telescopes at the La Silla Observatory under programme IDs 077.D-0092, 084.A-9016, and 085.A-9027. APEX is a collaboration between the Max-Planck-Institut f\"{u}r Radioastronomie, the European Southern Observatory, and the Onsala Space Observatory. This paper makes use of the following ALMA data: ADS/JAO.ALMA\#2011.0.00863.S and ADS/JAO.ALMA\#2011.0.00001.SV. ALMA is a partnership of ESO (representing its member states), NSF (USA) and NINS (Japan), together with NRC (Canada) and NSC and ASIAA (Taiwan), in cooperation with the Republic of Chile. The Joint ALMA Observatory is operated by ESO, AUI/NRAO, and NAOJ.
\end{acknowledgements}

\bibliographystyle{aa}
\bibliography{cdetect}

\begin{appendix}

\section{HD~100546 photospheric abundances}\label{sec:hd546abuns}

\begin{table}[!ht]
\centering
\caption{Photospheric abundances derived for HD~100546.}
\label{tab:hd546abuns}
\begin{tabular}{ c c c c c }
\hline
\hline
Element	&	Atomic	&	$\log_{10}{(\rm X/H)}_{\rm 546}$	& \#		& $\log_{10}{(\rm X/H)}_{\odot}$	\\
		&	number	&								& lines 	&		\\
\hline
He	&	2 &	$-1.00\pm0.12$	&   $*3$	& $-1.07\pm0.01$	\\
C	&	6 &	$-3.68\pm0.20$	&   $5$	& $-3.57\pm0.05$	\\
N	&	7 &	$-3.82\pm0.30$	&   $*1$	& $-4.17\pm0.05$	\\
O	&	8 &	$-3.25\pm0.10$	&   $5$	& $-3.31\pm0.05$	\\
Mg	&	12 &	$-5.41\pm0.12$	&   $5$	& $-4.40\pm0.04$	\\
Al	&	13 &	$-6.81\pm0.50$	&   $*1$	& $-5.55\pm0.03$	\\
Si	&	14 &	$-5.26\pm0.21$	&   $*2$	& $-4.49\pm0.03$	\\
S	&	16 &	$-5.37\pm0.50$	&   $*1$	& $-4.88\pm0.03$	\\
Ca	&	20 &	$-6.59\pm0.28$	&   $*3$	& $-5.66\pm0.04$	\\
Sc	&	21 &	$-10.18\pm0.40$	&   $*1$	& $-8.85\pm0.04$	\\
Ti	&	22 &	$-8.12\pm0.23$	&   $4$	& $-7.05\pm0.05$	\\
Cr	&	24 &	$-7.31\pm0.34$	&   $4$	& $-6.36\pm0.04$	\\
Fe	&	26 &	$-5.67\pm0.08$	&   $7$	& $-4.50\pm0.04$	\\
Ni	&	28 &	$-6.07\pm0.50$	&   $*2$	& $-5.78\pm0.04$	\\
Sr	&	38 &	$-10.01\pm0.30$	&   $*1$	& $-9.13\pm0.07$	\\
Ba	&	56 &	$-9.98\pm0.40$	&   $*2$	& $-9.82\pm0.09$	\\
\hline
\end{tabular}
\flushleft
\emph{Notes. }The number of lines used in the abundance fitting is shown in column 4. Asterisks denote very uncertain results. Solar abundances are from \citet{Asplundetal2009}.
\end{table}

The photospheric properties of HD~100546 were derived through directly fitting synthetic spectra to the optical spectrum of the star.  This process gives a measurements of Teff, $\log{(g)}$, vsini, and chemical abundances.  The synthetic spectra were calculated using the {\sc Zeeman} spectrum synthesis program \citep{Landstreet1988, Wadeetal2001}, using model stellar atmospheres from ATLAS9 \citep{Kurucz1993}, and atomic data from the Vienna Atomic Line Data Base \citep[VALD;][]{Kupkaetal1999}. The best-fit stellar parameters were derived using and automatic $\chi^{2}$ minimization routine, in which all parameters were fit simultaneously \citep{Folsometal2012}. 

This fitting process was performed on 9 independent spectral windows (416-427, 440-460, 460-480, 490-510, 510-520, 520-540, 540-560, 620-650, and 710-750~nm).  The final best fit parameters are the averages of the individual windows, and the uncertainties are the standard deviations.  In cases were a chemical abundance for an element was derived in only three or fewer windows, the uncertainties were increased to account for the full window-to-window variation, as well as any potential uncertainties due to line blending and continuum normalization.  Great care was taken to avoid any lines contaminated by emission.  Weak emission infilling of lines was identified by looking for lines with variability between the observations, and lines with asymmetries or anomalous profile shapes.  This fitting process is identical to that used by \citet{Folsometal2012}.  

The results of this fitting process were checked for consistence against the Balmer line profiles, and a good agreement was found.  However, Balmer lines were not used as the primary constraint on Teff or $\log{(g)}$, due to the uncertainties normalizing very broad lines in echelle spectra, and due to the emission in at least the core of the lines.

The stellar properties derived during the photospheric abundance analysis for HD~100546 are Teff = $10390 \pm 600$ K and logg = $4.20 \pm 0.30$. The projected rotational velocity is vsini = $64.9 \pm 2.2$ km/s.   Chemical abundances are presented in Table~\ref{tab:hd546abuns}. Microturbulence could not be well constrained for the star, although it is small ($< 2$ km/s), thus we assumed a value of $1$~km/s which is typical for a Herbig~Be star \citep[e.g.][]{Folsometal2012}.  An error of $1$~km/s in microturbulence would produce an error significantly less than $1\sigma$ in all our derived parameters. We find strong underabundances of iron-peak elements as well as Mg, Al, Si, Ca and Sc. However, we find abundances consistent with solar for the volatile elements He, C, N, and O. Chemically, this star appears to be a $\lambda$~Bo\"{o} star.

HD~100546 has parallax of $10.32 \pm 0.43$ mas \citep{vanLeeuwen2007} and a $V$ magnitude of $6.68 \pm 0.03$ \citep{Mermilliodetal1997, vanLeeuwen2007}. Using the bolometric correction of \citet{Balona1994} with our \teff, an intrinsic color from \citet{KenyonHartmann1995}, and a standard reddening law ($A_V = 3.1*E(B-V)$), we find a luminosity of \lstar~$=(25 \pm 7)\,$\lsol. With our \teff, the luminosity implies a radius of $(1.5 \pm 0.3)\,$\rsol.  Putting the star on the H-R diagram and comparing to the evolutionary tracks of \citet{Siessetal2000}, we find  \mstar~$=(2.3 \pm 0.2)\,$\msol. The H-R diagram position is consistent with the ZAMS or the end of the PMS, implying an age ${>}6\,$Myr, consistent with the estimate of \citet{vandenAnckeretal1997b}.

\section{Checking for extended emission around HD~100546}\label{sec:contaminationcheck}

\begin{figure}[!ht]
\includegraphics[clip=, width=1.0\linewidth]{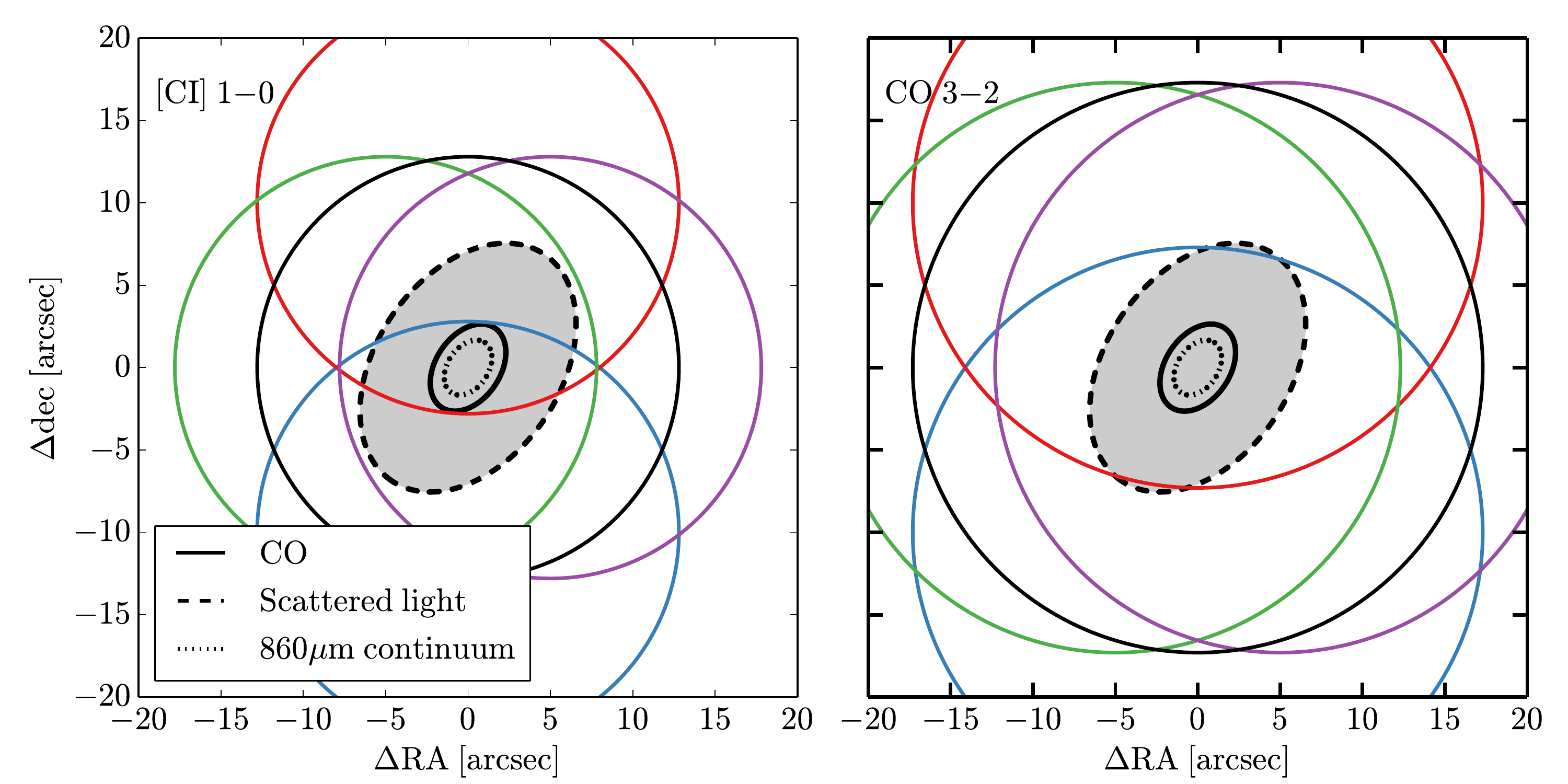}
\caption{The HD~100546 system. Black ellipses approximate the outer radius of CO~$3-2$ emission \citep[solid line, ][]{Walshetal2014b}, $\mu$m-sized grains as traced by scattered light \citep[dashed, ][]{Ardilaetal2007}, and the mm-sized grains as traced by the $860\mu$m continuum \citep[dotted, ][]{Walshetal2014b}. The APEX pointings from this work are shown as colored circles, corresponding directly to the spectra shown in Fig.~\ref{fig:hd100546spectra}.}
\label{fig:hd100546system}
\end{figure}

\begin{figure}[!ht]
\includegraphics[clip=, width=1.0\linewidth]{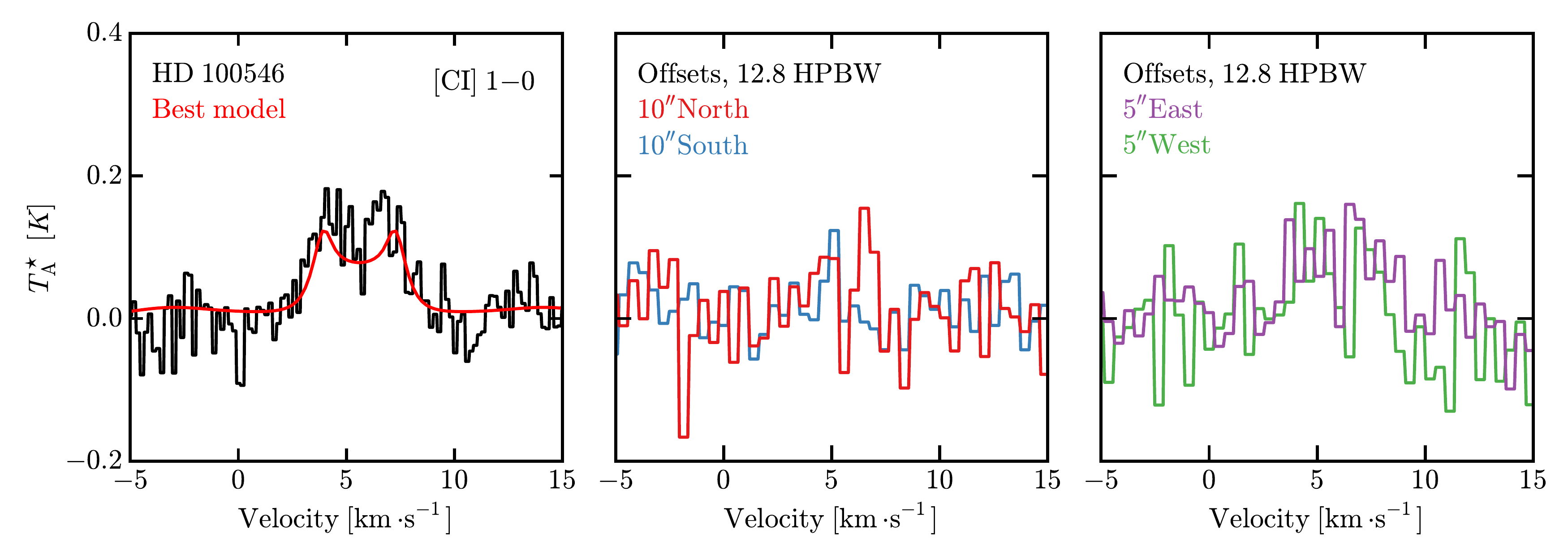}
\includegraphics[clip=, width=1.0\linewidth]{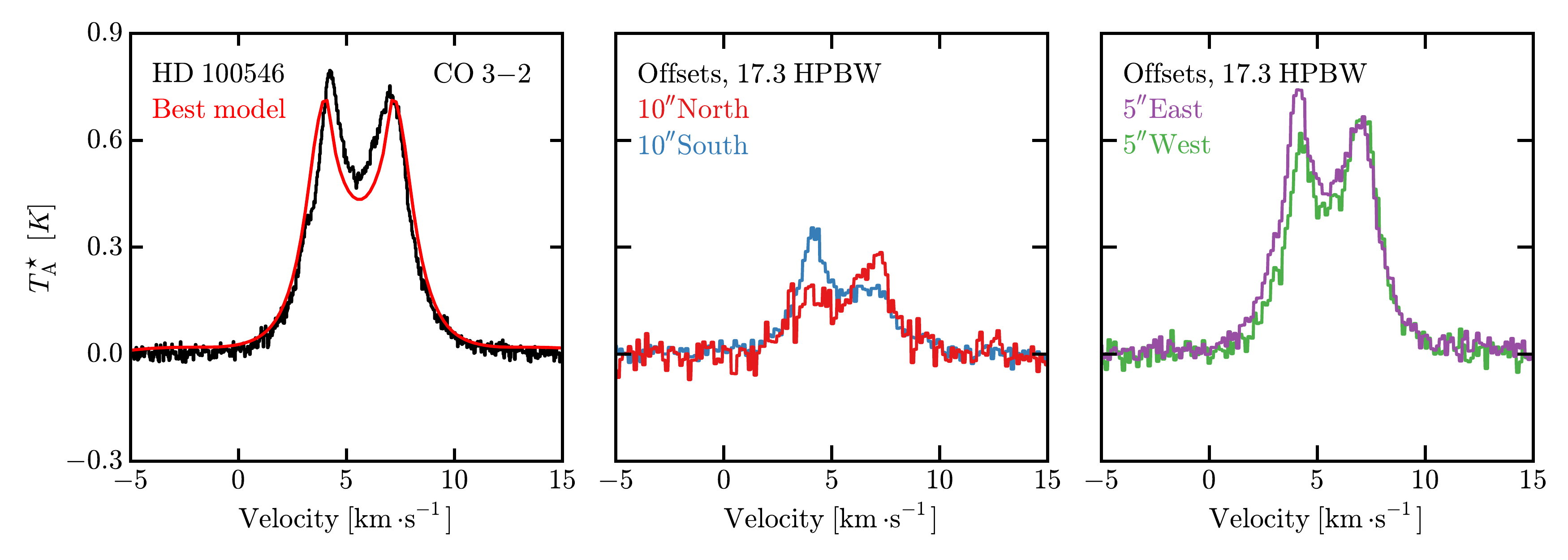}
\caption{Observations of the [CI]~$1-0$ and CO~$3-2$ transitions towards HD~100546 and four nearby reference positions (offset by $\pm5$'' East/West and $\pm10$'' North/South, as shown in Fig.~\ref{fig:hd100546system}). The emission strongly peaks at the disk and the red and blue peaks show variation expected from a Keplerian disk observed at the various offsets.}
\label{fig:hd100546spectra}
\end{figure}

A $\gtrsim 1000\,$au scale scattered light halo is seen around HD~100546 \citep{Gradyetal2001, Ardilaetal2007}, and the \cii\ emission is single-peaked, also suggesting a compact or extended envelope \citep{Fedeleetal2013a}. To determine the disk origin of the CO and \ci\ lines observed in this study, we performed four additional pointings on the \cionezero\ and the CO~$3$--$2$ lines (not discussed at length in the paper). These were offset by $\pm10''$ in the N-S direction, and by $\pm5''$ in the E-W direction. The pointings are shown in Fig.~\ref{fig:hd100546system}, and the resulting spectra are shown in Fig.~\ref{fig:hd100546spectra}. Considering the red- (N-E) and blueshifted (S-W) sides of the disk, all the spectra are fully consistent with the entire flux originating in the disk. We have also checked that the total CO~$3$--$2$ line flux in our APEX data is the same as that recovered from the ALMA observation of \citet{Walshetal2014b}. In summary, we confirm that the \cionezero\ and CO~$3$--$2$ transitions have no off-source contribution and that the CO line asymmetry represents a real azimuthal asymmetry in the disk emission, rather than pointing offsets.

\section{Summary of observational data}\label{apx:allobs}

In Table~\ref{tab:obsdata}, we list the line fluxes and upper limits used in the analysis. The spectral energy distribution (SED) for HD~100546 was adopted from \citet[][and references therein]{Muldersetal2011} and supplemented with ALMA measurements from \citet{Walshetal2014b}. The SED of TW~Hya was adopted from \citet[][and references therein]{Andrewsetal2012}.

\begin{table*}[!ht]
\centering
\caption{Line fluxes in units of $10^{-18}\,$W$\,$m$^{-2}$.}\label{tab:obsdata}
\begin{tabular}{ c c c c c}
\hline
\hline
Line	& HD 100546						&	ref.		& TW Hya			&	ref.	\\
\hline
CO $2$--$1$	&	(--)					&			&	$0.082\pm0.016$	&	$^{g}$	\\	
CO $3$--$2$	&	$1.72\pm0.01$		&	$^{a}$	&	$0.43\pm0.13$	&	$^{g}$	\\	
CO $6$--$5$	&	$16.1\pm0.8$			&	$^{a,b}$	&	$2.5\pm0.8$		&	$^{a,b}$	\\	
CO $7$--$6$	&	$23.5\pm2.7$			&	$^{c}$	&	(--)				&		\\	
CO $8$--$7$	&	$33.1\pm3.5$			&	$^{c}$	&	(--)				&		\\	
CO $9$--$8$	&	$45.3\pm4.1$			&	$^{c}$	&	(--)				&		\\	
CO $10$--$9$	&	$55.3\pm3.7$			&	$^{c}$	&	$3.4\pm0.6$		&	$^{h}$	\\	
CO $11$--$10$	&	$51.3\pm4.5$		&	$^{c}$	&	(--)				&		\\	
CO $12$--$11$	&	$58.3\pm3.3$		&	$^{c}$	&	(--)				&		\\	
CO $13$--$12$	&	$60.7\pm4.7$		&	$^{c}$	&	(--)				&		\\	
CO $14$--$13$	&	$60.0\pm8.3$		&	$^{d}$	&	(--)				&		\\	
CO $15$--$14$	&	$88.3\pm9.7$		&	$^{d}$	&	(--)				&		\\	
CO $16$--$15$	&	$58.8\pm9.7$		&	$^{d}$	&	(--)				&		\\	
CO $17$--$16$	&	$73.9\pm10.0$	&	$^{d}$	&	(--)				&		\\	
CO $18$--$17$	&	$71.5\pm6.9$		&	$^{d}$	&	$3.5\pm1.2$		&	$^{i}$	\\	
CO $19$--$18$	&	$64.7\pm8.5$		&	$^{d}$	&	(--)				&		\\	
CO $20$--$19$	&	$49.9\pm5.7$		&	$^{d}$	&	(--)				&		\\	
CO $21$--$20$	&	$65.0\pm8.8$		&	$^{d}$	&	(--)				&		\\	
CO $22$--$21$	&	$\leq42.0$		&	$^{d}$	&	(--)			&		\\	
CO $23$--$22$	&	$78.4\pm11.0$	&	$^{d}$	&	$\leq4.4$		&	$^{j}$	\\	
CO $24$--$23$	&	$71.3\pm12.0$	&	$^{d}$	&	(--)			&		\\	
CO $25$--$24$	&	$\leq77.1$		&	$^{d}$	&	(--)			&		\\	
CO $28$--$27$	&	$81.5\pm11.0$	&	$^{d}$	&	(--)			&		\\	
CO $29$--$28$	&	$78.6\pm19.0$	&	$^{d}$	&	$\leq3.4$		&	$^{i}$	\\	
CO $30$--$29$	&	$73.4\pm15.0$	&	$^{d}$	&	(--)			&		\\	
CO $31$--$30$	&	$\leq140.0$		&	$^{d}$	&	(--)			&		\\	
CO $32$--$31$	&	$\leq65.3$		&	$^{d}$	&	(--)			&		\\	
CO $33$--$32$	&	$\leq84.5$		&	$^{d}$	&	$\leq8.3$		&	$^{i}$	\\	
CO $34$--$33$	&	$53.1\pm14.0$	&	$^{d}$	&	(--)			&		\\	
CO $35$--$34$	&	$\leq44.1$		&	$^{d}$	&	(--)			&		\\	
CO $36$--$35$	&	$52.9\pm13.0$	&	$^{d}$	&	$\leq4.2$		&	$^{i}$	\\	
CO $37$--$36$	&	$\leq82.9$		&	$^{d}$	&	(--)			&		\\	
CO $38$--$37$	&	$\leq107.0$		&	$^{d}$	&	(--)			&		\\	
$^{13}$CO $2$--$1$	&	(--)			&			&	$0.020\pm0.001$	&	$^{k}$	\\	
$^{13}$CO $3$--$2$	&	$\leq0.66$	&	$^{e}$	&	$0.044\pm0.013$	&	$^{l}$	\\	
$^{13}$CO $6$--$5$	&	$\leq7.5$		&	$^{c}$	&	(--)			&	\\	
$^{13}$CO $7$--$6$	&	$\leq7.2$ 	&	$^{c}$	&	(--)			&	\\	
$^{13}$CO $8$--$7$	&	$\leq10.2$	&	$^{c}$	&	(--)			&	\\	
$^{13}$CO $9$--$8$	&	$\leq14.4$	&	$^{c}$	&	(--)			&	\\	
$^{13}$CO $10$--$9$	&	(--)			&			&	$0.28\pm0.04$	&	$^{h}$	\\	
$^{13}$CO $11$--$10$	& $16.8\pm3.4$	&	$^{c}$	&	(--)			&		\\	
$^{13}$CO $12$--$11$	&	(--)			&			&	(--)			&		\\
C$^{18}$O $2$--$1$	&	(--)			&			&	$0.006\pm0.001$	&	$^{k}$ \\	
HD $112\,\mu$m	&	$\leq27.0$		&	$^{a}$	&	$6.3\pm0.7$	&	$^{j}$	\\	
HD $56\,\mu$m	&	$\leq160.0$		&	$^{a}$	&	$3.1\pm1.0$	&	$^{n}$	\\	
CI $1$--$0$		&	$0.66\pm0.07$	&	$^{a,b}$	&	$0.067\pm0.018$	&	$^{a,b}$	\\	
CI $2$--$1$		&	$\leq3.58$		&	$^{a}$	&	$\leq0.9$		&	$^{a}$	\\	
CII $158\,\mu$m	&	$135.0\pm15.0$	&	$^{d,f}$	&	$\leq6.6$		&	$^{m}$	\\	
OI $145\,\mu$m	&	$357.0\pm13.0$	&	$^{d,f}$	&	$\leq5.5$		&	$^{m}$	\\	
OI $63\,\mu$m	&	$5540.0\pm50.0$	&	$^{d,e}$	&	$36.5\pm12.1$	&	$^{m}$	\\	
C$_{2}$H\,$N{=}4{-}3,J{=}\frac{9}{2}{-}\frac{7}{2},F{=}5{-}4$	&	$\leq0.005$		&	$^{a}$	&	$0.017\pm0.008$	&	$^{a}$	\\	
C$_{2}$H\,$N{=}4{-}3,J{=}\frac{9}{2}{-}\frac{7}{2},F{=}4{-}3$	&	$\leq0.005$		&	$^{a}$	&	$0.020\pm0.008$	&	$^{a}$	\\	
C$_{2}$H\,$N{=}4{-}3,J{=}\frac{7}{2}{-}\frac{5}{2},F{=}4{-}3$	&	$\leq0.005$		&	$^{a}$	&	$0.019\pm0.008$	&	$^{a}$	\\	
C$_{2}$H\,$N{=}4{-}3,J{=}\frac{7}{2}{-}\frac{5}{2},F{=}3{-}2$	&	$\leq0.005$		&	$^{a}$	&	$0.024\pm0.008$	&	$^{a}$	\\	
\hline
\end{tabular}
\flushleft
\emph{References. }$^{a}$ -- this work; $^{b}$ -- \citet{Kamaetal2016a}; $^{c}$ -- \citet{vanderWieletal2014}; $^{d}$ -- \citet{Meeusetal2013}; $^{e}$ -- \citet{Panicetal2010}; $^{f}$ -- \citet{Fedeleetal2013a}; $^{g}$ -- \citet{Qietal2004}; $^{h}$ -- Fedele et al. (submitted); $^{i}$ -- \citet{Kampetal2013}; $^{j}$ -- \citet{Berginetal2013}; $^{k}$ -- \citet{Favreetal2013}; $^{l}$ -- \citet{Thietal2004}; $^{m}$ -- \citet{Thietal2010}; $^{n}$ -- \citet{Fedeleetalinprep}.
\end{table*}

\section{Abundance and emission maps}\label{apx:linemaps}

\begin{figure*}[!ht]
\includegraphics[clip=,width=1.0\linewidth]{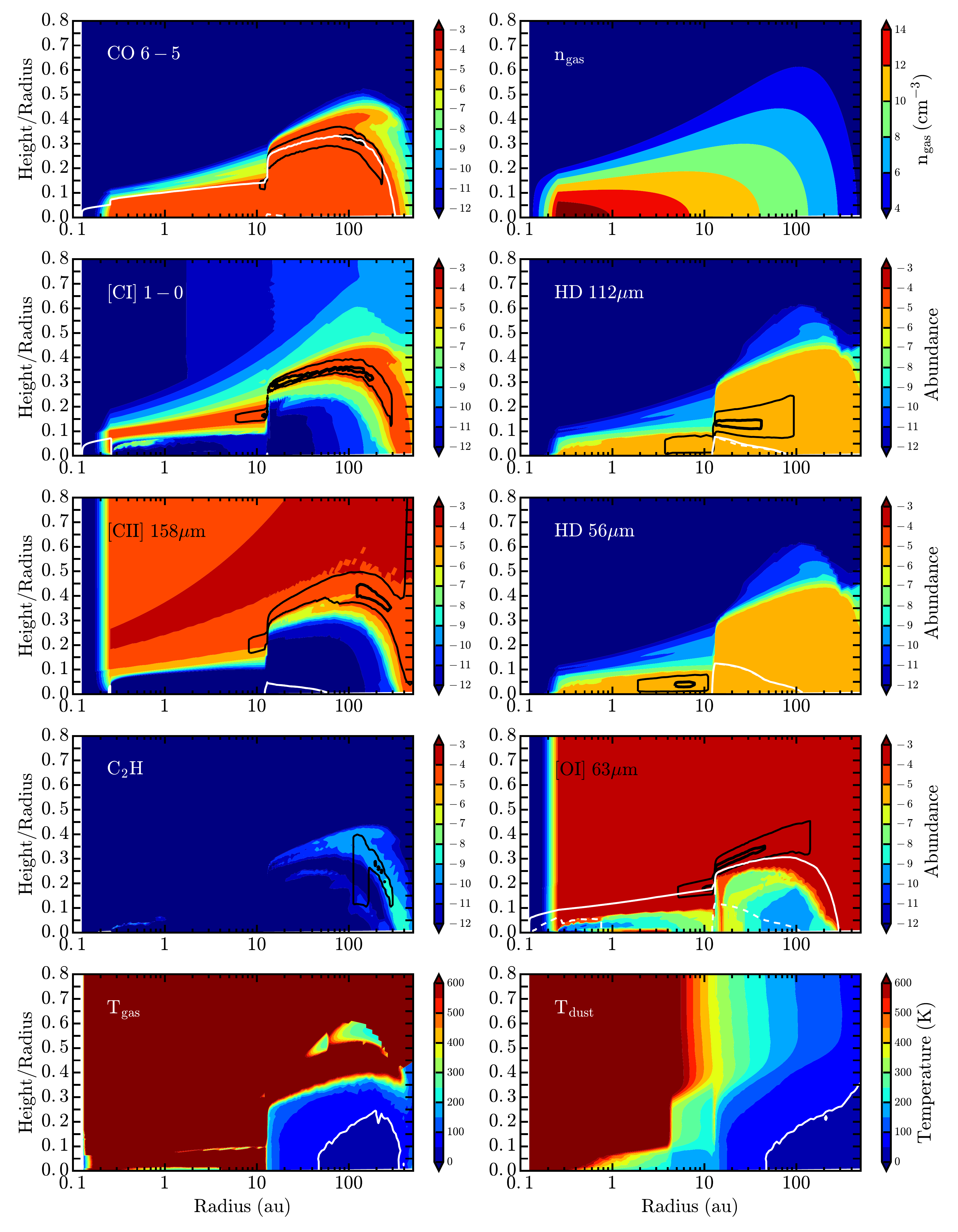}
\caption{The disk structure and line contribution functions (CBFs) for the HD~100546 model with \gdrat${=}10$ and \chgas${=}1.35\times10^{-4}$ (ISM). The CBF panels show the abundance (colour fill) and line emission (black lines) of CO, \ci, C$_{2}$H, \cii, and HD. The black contours encompass $25$ and $75\,$\% cumulative contributions to the line emission. The solid and dashed white lines show the $\tau=1$ surface for the line and continuum emission. The solid white line for $T_{\rm gas}$ and $T_{\rm dust}$ is the $50\,$K isotherm.}
\label{fig:hd100546cbfs}
\end{figure*}

\begin{figure*}[!ht]
\includegraphics[clip=,width=1.0\linewidth]{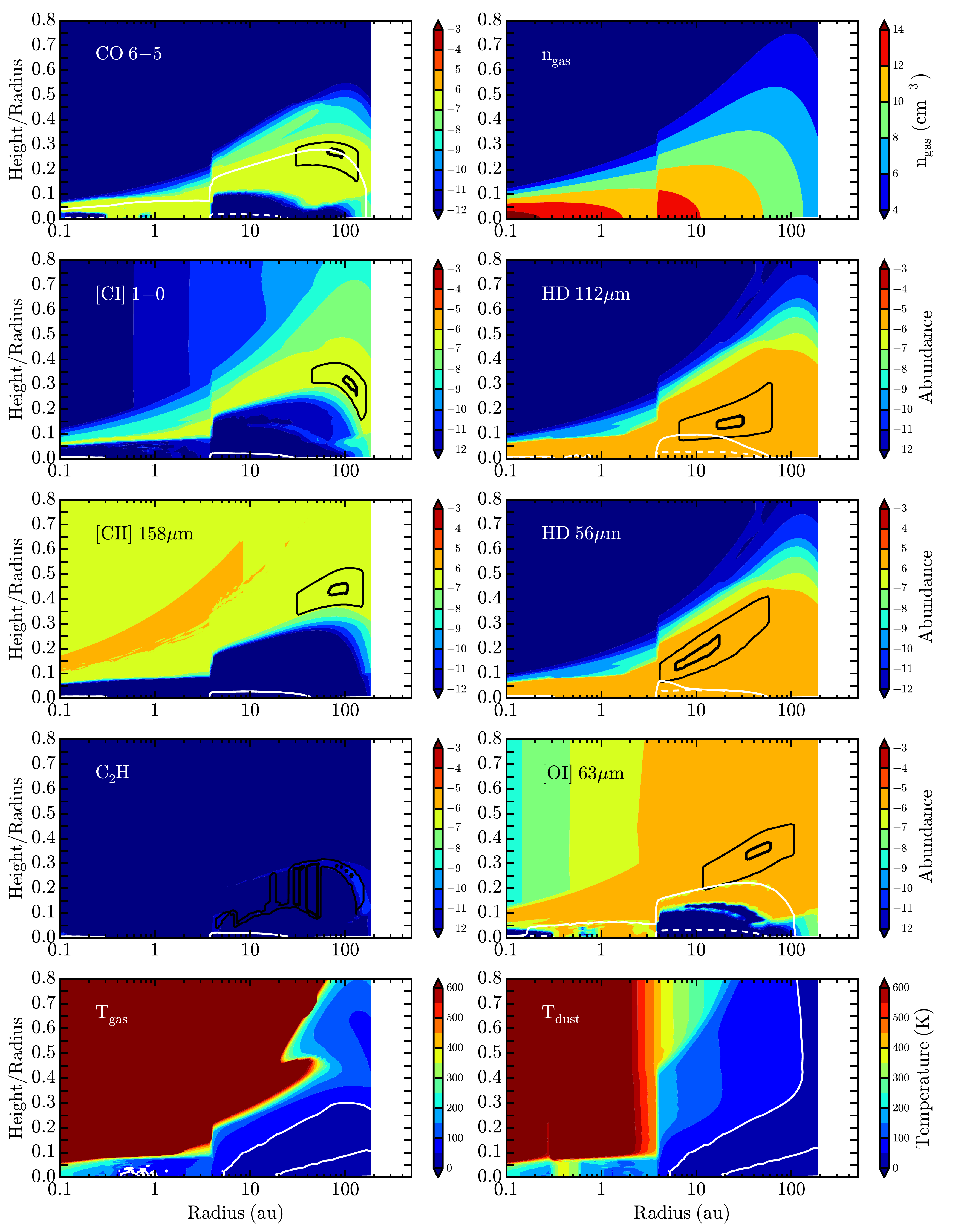}
\caption{The disk structure and line contribution functions (CBFs) for the TW~Hya model with  \gdrat${=}200$ and \chgas${=}10^{-6}$ (depleted). The CBF panels show the abundance (colour fill) and line emission (black lines) of CO, \ci, C$_{2}$H, \cii, and HD. The black contours encompass $25$ and $75\,$\% cumulative contributions to the line emission. The solid and dashed white lines show the $\tau=1$ surface for the line and continuum emission. The solid white lines for $T_{\rm gas}$ and $T_{\rm dust}$ are the $20$ and $50\,$K isotherms.}
\label{fig:twhyacbfs}
\end{figure*}

In Figs.~\ref{fig:hd100546cbfs} and \ref{fig:twhyacbfs}, we show the gas density, gas and dust temperature, and abundances of species relevant to this work in the HD~100546 and TW~Hya disk. The main parameter in each panel is shown as a colormap. The abundances are overlaid in black with the $25$ and $75\,$\% contours of the line contribution function. The contribution function represents the cumulative buildup of line emission, vertically integrated from top to bottom, and radially outward from the grid cell closest to the star at each height. The plotted contours show the area where both the vertical and radial cumulative emission contribution function for a given line is between $12.5$ and $87.5\,$\% (the $25\,\%$ contour), or between $37.5$ and $62.5\,$\% (the $75\,$\% contour).

\end{appendix}
   
\end{document}